\documentclass[letter,traditabstract]{aa} % for the letters 

\usepackage{graphicx}
\usepackage{txfonts}
\usepackage{longtable}
\begin{document}

   \title{A Virgo Environmental Survey Tracing Ionised Gas Emission (VESTIGE).VI. Environmental quenching on HII region scales \thanks{Based on observations obtained with
   MegaPrime/MegaCam, a joint project of CFHT and CEA/DAPNIA, at the Canada-French-Hawaii Telescope
   (CFHT) which is operated by the National Research Council (NRC) of Canada, the Institut National
   des Sciences de l'Univers of the Centre National de la Recherche Scientifique (CNRS) of France and
   the University of Hawaii. }
      }
   \subtitle{}
  \author{A. Boselli\inst{1},  
          M. Fossati\inst{2},
	  A. Longobardi\inst{1},
	  %G. Consolandi\inst{},
	  %P. Amram\inst{1},
	  %M. Sun\inst{6},
	  %P. Andreani\inst{7},  
	  S. Boissier\inst{1},
	  M. Boquien\inst{3},
	  J. Braine\inst{4},
	  %F. Combes\inst{10,11},
          P. C{\^o}t{\'e}\inst{5},
          J.C. Cuillandre\inst{6},
	  %P.A. Duc\inst{14},
	  %E. Emsellem\inst{7},
	  B. Epinat\inst{1},
          L. Ferrarese\inst{5},
	  G. Gavazzi\inst{7},
          S. Gwyn\inst{5}, 
	  G. Hensler\inst{8},
	  %E.W. Peng\inst{17,4},
	  H. Plana\inst{9},
	  %J. Roediger\inst{12},
	  Y. Roehlly\inst{1},
	  %R. Sanchez-Janssen\inst{19},
	  %M. Sarzi\inst{20},
	  C. Schimd\inst{1},
	  %P. Serra\inst{21},
	  M. Sun\inst{10},
	  G. Trinchieri\inst{11}  
       }

\institute{     
                Aix Marseille Univ, CNRS, CNES, LAM, Marseille, France
                \email{alessandro.boselli@lam.fr, alessia.longobardi@lam.fr}
        \and
                Institute for Computational Cosmology and Centre for Extragalactic Astronomy, Department of Physics, Durham University, South Road, Durham DH1 3LE, UK
		\email{matteo.fossati@durham.ac.uk}
        %\and
	%	Kavli Institute for Astronomy and Astrophysics, Peking University, Beijing 100871, PR China	
        %\and
	%	Department of Physics and Astronomy, University of Alabama in Huntsville, Huntsville, AL 35899, USA
 	%\and
 	%	European Southern Observatory, Karl-Schwarzschild-Strasse 2, 85748, Garching, Germany 
	\and
		Centro de Astronom\'a (CITEVA), Universidad de Antofagasta, Avenida Angamos 601, Antofagasta, Chile
	\and
		Laboratoire d'Astrophysique de Bordeaux, Univ. Bordeaux, CNRS, B18N, all\'ee Geoffroy Saint-Hilaire, 33615 Pessac, France
	%\and	
	%	College de France, 11 Pl. M. Berthelot, F-75005 Paris, France 
        %\and
        %        LERMA, Observatoire de Paris, CNRS, PSL Research University, Sorbonne Universit\'es, UPMC Univ. Paris 06, F-75014 Paris, France
        \and  
                NRC Herzberg Astronomy and Astrophysics, 5071 West Saanich Road, Victoria, BC, V9E 2E7, Canada
        \and        
		AIM, CEA, CNRS, Universit\'e Paris-Saclay, Universit\'e Paris Diderot, Sorbonne Paris Cit\'e, Observatoire de Paris, PSL University, F-91191 Gif-sur-Yvette Cedex, France
	%\and
	%	Observatoire Astronomique de Strasbourg, UMR 7750, 11, rue de l'Universit\'e, 67000, Strasbourg, France
	\and
		Universit\'a di Milano-Bicocca, piazza della scienza 3, 20100 Milano, Italy
	\and
		Department of Astrophysics, University of Vienna, T\"urkenschanzstrasse 17, 1180 Vienna, Austria
	%\and
	%	Department of Astronomy, Peking University, Beijing 100871, PR China
	\and
		Laborat\'orio de Astrof\'isica Te\'orica e Observacional, Universidade Estadual de Santa Cruz - 45650-000, Ilh\'eus-BA, Brasil
	%\and
	%	UK Astronomy Technology Centre, Royal Observatory Edinburgh, Blackford Hill, Edinburgh, EH9 3HJ, UK
	%\and
	%	Centre for Astrophysics Research, University of Hertfordshire, Hatfield AL10 9AB, UK
	%\and
	%	Osservatorio Astronomico di Cagliari, via della scienza 5, 09047 Selargius, Cagliari, Italy
	\and
		Department of Physics and Astronomy, University of Alabama in Huntsville, Huntsville, AL 35899, USA
	\and
		INAF - Osservatorio Astronomico di Brera, via Brera 28, 20159 Milano, Italy
                 }

\authorrunning{Boselli et al.}
\titlerunning{VESTIGE}

   \date{}

% \abstract{}{}{}{}{} 
% 5 {} token are mandatory
 
  \abstract  
{The Virgo Environmental Survey Tracing Ionised Gas Emission (VESTIGE) is a blind narrow-band 
H$\alpha$+[NII] imaging survey of the Virgo cluster carried out with MegaCam at the Canada-French-Hawaii telescope (CFHT).
We use a new set of data extracted from VESTIGE to study the impact of the 
hostile cluster environment on the star formation process down to the scale of HII regions ($\sim$ 50 pc). HII regions are identified
and their parameters measured using the \textsc{HIIphot} code on a sample of 114 late-type galaxies spanning a wide range in morphological type (Sa-Sd, Im, BCD),
stellar mass (10$^{6.5}$ $\leq$ $M_{star}$ $\leq$ 10$^{11}$ M$_{\odot}$), and star formation activity (10$^{-3}$ $\leq$ $SFR$ $\leq$ 10 M$_{\odot}$ yr$^{-1}$). 
Owing to the exquisite average resolution of the VESTIGE data (0.65 arcsec), we detect 11302 HII regions with an H$\alpha$ luminosity $L(H\alpha)$ $\geq$ 10$^{37}$ erg s$^{-1}$.
We show that the typical number of HII regions in gas-stripped
objects is significantly lower than in healthy late-types of similar stellar mass. We also show that in these gas-stripped galaxies the number of HII regions
significantly drops outside the effective radius, suggesting that the quenching process occurs outside-in, in agreement with other multifrequency observations. 
These new results consistently confirm that the main mechanism responsible for the decrease of the star formation activity observed in cluster galaxies is ram pressure,
allowing us to discard other milder processes such as starvation or strangulation unable to reproduce the observed radially truncated profiles.
  
 }
   {}
   {}
   {}
   {}
   {}


   \maketitle
%
%________________________________________________________________

\section{Introduction}

Environment plays a major role in shaping galaxy evolution. Clusters of galaxies are mainly composed of
early-type systems, while the field is dominated by star forming spirals and irregulars
(Dressler 1980). It is also well established that late-type galaxies in clusters are characterised 
by a lower atomic gas (Cayatte et al. 1990; Solanes et al. 2001; Gavazzi et al. 2005; Chung et al. 2009), molecular gas
(Fumagalli et al. 2009, Boselli et al. 2014a), and dust content (Cortese et al. 2010, 2012a)
than similar objects in the field. Because they lack gas in its different phases, cluster galaxies also have reduced
star formation activity (Lewis et al. 2002; Gomez et al. 2003; Gavazzi et al. 2006; Boselli et al. 2015, 2016). Different mechanisms 
have been proposed to explain the relative differences in evolution between cluster members and isolated galaxies, 
as reviewed in Boselli \& Gavazzi (2006, 2014). These can be divided into two main families: 
the gravitational perturbations induced on a galaxy during its interaction with other cluster members or with the potential
well of the cluster (tidal interactions, harassment - Merritt 1983, Byrd \& Valtonen 1990, Moore et al. 1998),
and the dynamical interaction of the galactic ISM with the hot and dense intracluster medium trapped within the potential well 
of the cluster (ram pressure stripping, thermal evaporation, starvation - Gunn \& Gott 1972, Cowie \& Songaila 1977, Larson et al. 1980).
The identification of the dominant mechanism in high density regions, like clusters and groups, at different epochs is still under debate 
(Dressler et al. 1997, Poggianti et al. 1999, 2017, Dressler 2004, De Lucia et al. 2012, Muzzin et al. 2014, Boselli et al. 2019a).

These perturbing mechanisms have different effects on the star formation activity and on the quenching 
process in perturbed galaxies. For example, gravitational perturbations generally induce gas infall in the nuclear regions, 
thus are expected to increase the activity of star formation toward the nucleus and in the turbulent regions formed 
during the interaction (Keel et al. 1985, Sun et al. 2007, Patton et al. 2011). On the contrary, milder processes such as the reduced infall of freshly 
accreted gas all over the disc would uniformly decrease the star formation activity at all galactocentric distances on long time-scales ($\gtrsim$ 5 Gyr, starvation; Boselli et al. 2006).
Finally, ram-pressure stripping, responsible for an outside-in removal of the gas, would produce truncated gaseous and 
star forming discs on shorter time-scales ($\lesssim$ 500 Myr; Vollmer et al. 2004, 2006; Boselli et al. 2006, 2016; Crowl \& Kenney 2008; 
Gullieuszik et al. 2017; Fossati et al. 2018). The effect 
of these different mechanisms on smaller scales, those of the collapsing giant molecular clouds and of the HII regions
where stars are formed, is however totally unknown. A few nearby perturbed galaxies have been studied in detail at high angular resolution with HST.
These works have produced some notable results such as the discovery of compact ($\sim$ 20 pc), young star clusters formed in the perturbed regions of the Antennae (Whitmore et al. 1995), 
or the presence of filamentary dust structures with associated star forming regions of size $\sim$ 100 pc
produced during the stripping of the ISM in the Coma cluster spirals NGC 4921 (Kenney et al. 2015) and D100 (Cramer et al. 2019).
The major limitation of these works, however, is that they are limited to a few peculiar objects. There is, indeed, an
objective difficulty in observing with a sufficient angular resolution and sensitivity a statistically significant 
sample of cluster galaxies. For instance, the Virgo cluster, which is the closest massive 
cluster of galaxies ($M_{cluster}$ $\simeq$ 4 $\times$ 10$^{14}$ M$_{\odot}$, McLaughlin 1999), is 
located at 16.5 Mpc (Gavazzi et al. 1999, Mei et al. 2007, Blakeslee et al. 2009); thus for the typical seeing of 2 arcsec 
of narrow-band H$\alpha$ imaging data available on the net (e.g. GOLDmine, Gavazzi et al. 2003),
the linear resolution is limited to $\sim$ 150 pc, insufficient to resolve individual HII regions in crowded regions such as spiral arms. 

The Virgo Environmental Survey Tracing Ionised Gas Emission (VESTIGE, Boselli et al. 2018) is a new deep blind
H$\alpha$ narrow-band (NB) imaging survey of the Virgo cluster carried out at the CFHT with MegaCam. Designed to cover 
the whole Virgo cluster up to its virial radius (104 deg$^2$), this survey is perfectly suited to study the 
effects of the environment on the star formation process in perturbed galaxies down to scales of $\lesssim$ 100 pc. Indeed, its blind nature secures
the observation of galaxies under the effect of any kind of perturbation, and the use of a NB filter centred on the H$\alpha$ 
emission line, whose emission in star forming galaxies is largely dominated by the radiation of the HII regions surrounding 
young ($t$ $\leq$ 10 Myr) massive ($M$ $\geq$ 8 M$_{\odot}$) stars, is an ideal tracer of star formation 
(e.g. Kennicutt 1998, Boselli et al. 2009). The depth of the survey and the extraordinary high imaging quality
make VESTIGE a unique instrument for this purpose. Although the survey is ongoing, 
the analysis done so far on a representative subsample of objects allows us to make the
first statistical study of the environmental quenching process down to the scale of individual HII regions.
This letter is aimed at presenting these spectacular results. A complete analysis of the full sample, including the study
of the typical scaling relations of HII regions in perturbed and unperturbed galaxies, will be published once the survey will be completed.
IFU spectroscopic observations, now mainly limited to unresolved star forming complexes in the tails of stripped 
material (Fossati et al. 2016, Poggianti et al. 2019), will provide a complementary information for understanding 
the star formation process in perturbed systems.

\section{Observations and data reduction}

\subsection{Narrow-band imaging data}

The VESTIGE observations are carried out using the NB filter 
MP9603 ($\lambda_c$ = 6590 \AA; $\Delta\lambda$ = 104 \AA) which is optimised for the 
Virgo cluster galaxies with a typical recessional velocity of 
-200 $\lesssim$ $cz$ $\lesssim$ 3000 km s$^{-1}$. At this redshift the filter
includes the H$\alpha$ line and the [NII] doublet.\footnote{Hereafter we refer to the H$\alpha$+[NII] 
band simply as H$\alpha$, unless otherwise stated.} The NB filter has a very flat 
transmissivity profile ($T$ $\simeq$ 92\%).
A detailed description of the observing strategy and of the data reduction procedure is
given in Boselli et al. (2018). %Here we summarise the information most relevant 
%for the following analysis.
%The full depth of the survey is reached through twelve independent exposures of 10 min
%integration each in the narrow-band filter, and 1 min in the broad-band $r$ filter, with this last 
%step being necessary for the subtraction of the stellar continuum. A large dither is used to 
%minimise any large-scale gradient in the sky background. The pixel scale is
%0.187 arcsec pixel$^{-1}$. The sensitivity of the survey is 
%$f(H\alpha)$ $\sim$ 4 $\times$ 10$^{-17}$ erg s$^{-1}$ cm $^{-2}$ for point sources (5 $\sigma$),
%while the median seeing is 0.65 arcsec.
At present, the survey is $\sim$ 36\% ~ complete, with the full depth (where the sensitivity is 
$f(H\alpha)$ $\sim$ 4 $\times$ 10$^{-17}$ erg s$^{-1}$ cm $^{-2}$ for point sources (5 $\sigma$)) reached in the central 
$\sim$ 16 deg$^2$ around M87, and in four independent outer regions $\sim$ 1 deg$^2$ each mapped during pilot 
observations. 
Shallower exposures are also available in an annulus of $\sim$ 20 deg$^2$ 
located around the central region. This low completion rate is due to very poor weather conditions at Mauna Kea
in 2018 and 2019. The median seeing is 0.65 arcsec. NB imaging data are at present available for
171 spectroscopically confirmed late-type Virgo cluster members catalogued in the Virgo Cluster Catalogue (VCC, Binggeli et al. 1985).

\subsection{HII regions identification}

HII regions are identified using the \textsc{HIIphot} data reduction pipeline (Thilker et al. 2000),
a code specifically designed to measure the photometric properties of HII regions from
NB imaging data. Owing to its quality this code became the reference in the literature 
for this kind of studies (e.g. Scoville et al. 2001; Helmbolt et al. 2005; Azimul et al. 2011; Lee et al. 2011; Liu et al. 2013). 
The code runs on the H$\alpha$ continuum subtracted, on the H$\alpha$
NB, and on the stellar-continuum images and uses a recognition technique through
an iterative growing procedure to identify single HII regions and measure their parameters.
The stellar continuum image is derived as in Boselli et al. (2019b) from the $r$-band image using a colour correction 
to take into account the dependence on the spectra shape of the emitting sources (Spector et al. 2012).
As extensively described in previous works (e.g. Pleuss et al. 2000; Bradley et al. 2006), 
the major limitation of \textsc{HIIphot} is that
of deriving accurate parameters in crowded regions whenever the seeing is poor and the 
sensitivity limited. Thanks to the excellent quality of the VESTIGE data
we can measure the photometric parameters of HII
regions down to $L(H\alpha)$ $\simeq$ 10$^{36}$ erg s$^{-1}$ and effective radii\footnote{$r(H\alpha)$ are the radii of the circles of surface 
equivalent to the area of the HII regions detected by \textsc{HIIphot}.} of 
$r(H\alpha)$ $\simeq$ 25 pc at the typical distance of the Virgo cluster (16.5 Mpc).
To avoid possible systematic effects at these low emission levels, where 
incompleteness could be important, we limit the present analysis
to the HII regions with an H$\alpha$ luminosity $L(H\alpha)$ $\geq$ 10$^{37}$ erg s$^{-1}$.
For these regions, the typical signal-to-noise is $S/N$ $>$ 100 with no dependence on 
the position of the source with respect to the galactocentric distance. 
Incompleteness should thus be marginal. We recall, however, that 
this work is only based on a comparative analysis of gas-rich and HI-stripped cluster galaxies,
for which the same limiting H$\alpha$ luminosity in the identification of the HII regions 
is applied, further minimising systematic effects.

We run the \textsc{HIIphot} code on all the late-type galaxies covered during the VESTIGE survey
and showing a clumpy emission in the continuum-subtracted H$\alpha$ frame witnessing the presence of star forming regions.
To limit possible selection biases, however, we restrict the present analysis to 
objects with sufficiently deep imaging data, i.e. with at least four independent exposures 
(40 minutes of integration in the NB). We also avoid highly inclined galaxies (axis ratio $b/a$ $<$ 0.35)
where projection effects or a strong attenuation could induce systematic effects in the 
determination of the physical parameters of the HII regions. These galaxies span a wide range in
morphological type (Sa-Sd, Im, BCD), stellar mass (10$^{6.5}$ $\leq$ $M_{star}$ $\leq$ 10$^{11}$ M$_{\odot}$), with a dominance of 
dwarf systems (62\% with $M_{star}$ $\leq$ 10$^{9}$ M$_{\odot}$, 10\% with $M_{star}$ $>$ 10$^{10}$ M$_{\odot}$),
and star formation rate (10$^{-3}$ $\leq$ $SFR$ $\leq$ 10 M$_{\odot}$ yr$^{-1}$).
To avoid any possible contamination from background line emitters ([OII], [OIII], H$\beta$), the extraction of HII regions 
is limited within the optical disc of the galaxy.

To identify objects which have been stripped of their gas in the hostile cluster environment we use the HI-deficiency parameter,
defined as the difference in logarithmic scale between the expected and the observed
HI mass of a galaxy of a given angular size and morphological type (Haynes \& Giovanelli 1984).
Being the baryonic component least bound to the gravitational potential well of the galaxy,
the atomic gas is easily removed in any kind of interaction (Boselli \& Gavazzi 2006). 
As defined, the HI-deficiency parameter is a normalised entity and can be used to quantify the degree of stripping 
that a galaxy is suffering in the hostile cluster environment. Indeed, the mean HI-deficiency of Virgo cluster members 
is known to increase towards the inner regions of the cluster (Cayatte et al. 1990, Solanes et al. 2001, Chung et al. 2009, Gavazzi et al. 2013, Boselli et al. 2014b).  
Consistently with Boselli et al. (2014b), the HI-deficiency of these VCC galaxies has been derived using the HI data 
available in the GOLDmine database (Gavazzi et al. 2003) using the morphology-dependent calibrations of Boselli \& Gavazzi (2009).
To have a similar number of normal (gas-rich) and stripped (gas-poor) objects we adopt $HI-def$=0.5 as threshold
in the HI-deficiency parameter. This relatively high threshold is 
indicative of galaxies with a quite low gas content given that the typical dispersion of
the HI-deficiency parameter in isolated galaxies is just 0.3.
According to these criteria, the total number of galaxies analysed in this work is 114 with
11302 HII regions, divided into 49 gas-rich and 65 gas-deficient objects with 7644 and 3658 HII regions each.
Consistent with Boselli et al. (2014b), stellar masses are derived using the Zibetti et al. (2009)
$g-i$ colour-dependent relation and a Chabrier IMF, with $g$ and $i$ magnitudes and $r$-band 
effective radii taken from Cortese et al. (2012b) when available, otherwise from the EVCC
(Kim et al. 2014). Distances for galaxies belonging to the different
Virgo cluster subgroups are all taken from Gavazzi et al. (1999), but
Virgo cluster A which is assumed at 16.5 Mpc. NB imaging fluxes
are corrected for [NII] contamination using the mean values derived from drift scan spectroscopy
whenever available (Boselli et al. 2015, Gavazzi et al. 2004, in order of priority), 
otherwise estimated from the mean [NII]/H$\alpha$ vs. $M_{star}$ scaling relation given in Boselli et al. (2009).

\section{Analysis} 

   \begin{figure}
   \centering
   \includegraphics[width=0.43\textwidth]{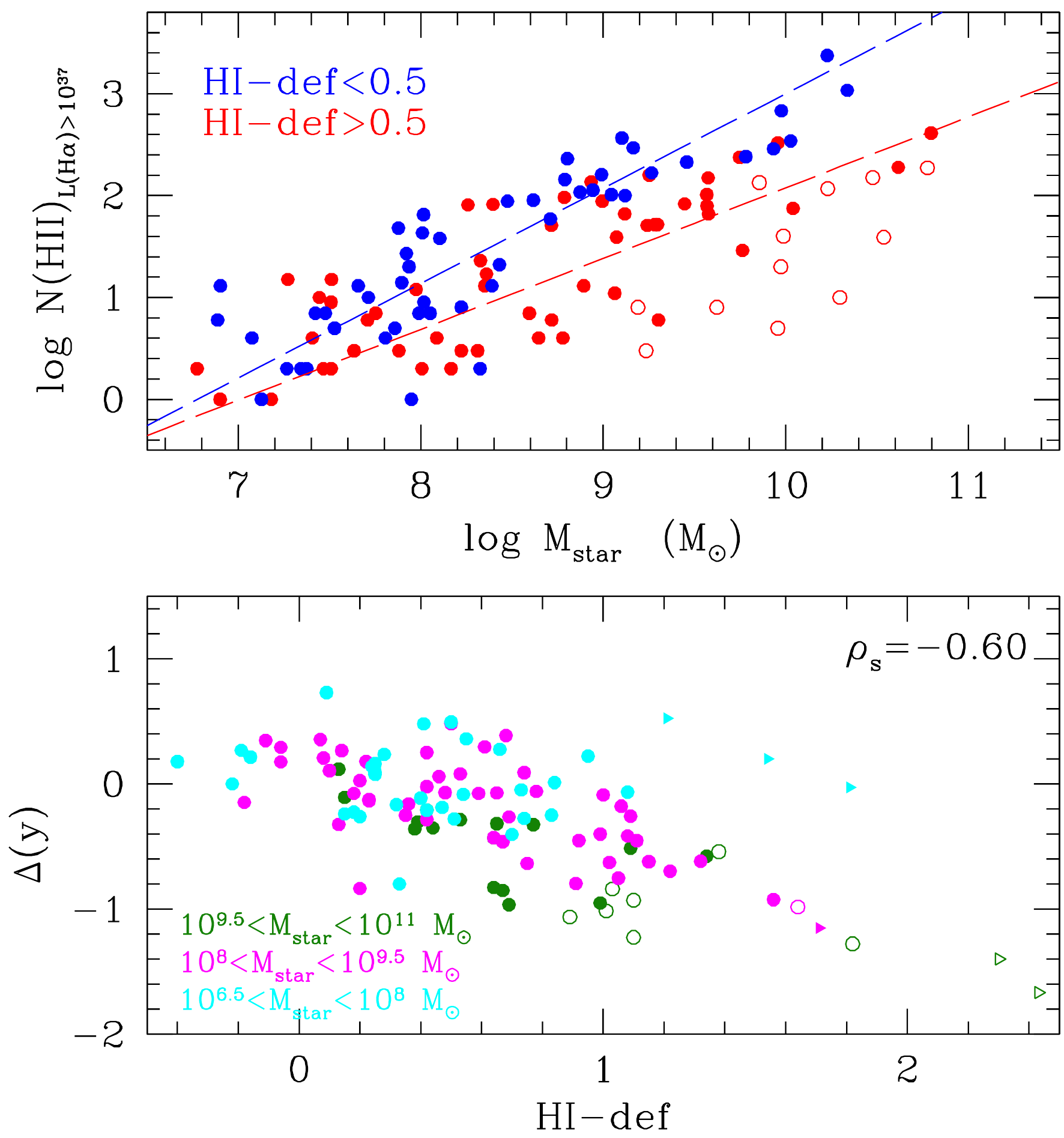}
   \caption{Upper panel: relationship between the number of HII regions of luminosity $L(H\alpha)$ $>$ 10$^{37}$ erg s$^{-1}$ and the stellar mass of galaxies.
   Blue symbols indicate gas-rich systems ($HI-def$ $\leq$ 0.5), red symbols HI-stripped galaxies ($HI-def$ $>$ 0.5).
   Empty symbols are for bulge-dominated Sa-Sab galaxies. The dashed blue and red lines give the bisector fits for gas-normal and gas-stripped systems.
   Lower panel: relationship between the residual of the above relation derived for HI-rich galaxies 
   and the HI-deficiency parameter. Symbols are colour-coded in ranges of stellar mass: 
   $M_{star}$ $>$ 10$^{9.5}$ M${_\odot}$ - dark green, 10$^{8}$ $<$ $M_{star}$ $\leq$ 10$^{9.5}$ M${_\odot}$ - magenta, 
   10$^{6.5}$ $<$ $M_{star}$ $\leq$ 10$^{8}$ M${_\odot}$ - cyan. Dots are for HI-detected galaxies, triangles for HI-undetected objects 
   (lower limits in the HI-deficiency parameter). $\rho_s$ is the Spearman correlation coefficient. 
   }
   \label{scalingmass}%
   \end{figure}

Figure \ref{scalingmass} shows the relationship between the number of HII regions of luminosity $L(H\alpha)$ $>$ 10$^{37}$ erg s$^{-1}$ and the stellar mass of galaxies.
Galaxies are divided into gas-rich objects (blue symbols) and gas-poor, stripped systems (red symbols) according to their HI-deficiency parameter.
Figure \ref{scalingmass} shows that, in systems with a normal gas content ($HI-def$ $\leq$ 0.5), the number of HII regions of luminosity $L(H\alpha)$ $>$ 10$^{37}$ erg s$^{-1}$
increases from $N(HII)$ $\lesssim$ 10 in galaxies of stellar mass $M_{star}$ $\simeq$ 10$^7$ M$_{\odot}$ to $N(HII)$ $\simeq$ 10$^3$ in spirals of $M_{star}$  
$\simeq$ 10$^{10}$ M$_{\odot}$. A flatter relation is followed by HI-poor galaxies, which have a similar number of HII regions compared to gas-rich
systems at low stellar masses but a significantly lower number of star forming regions at $M_{star}$  $\simeq$ 10$^{10}$ M$_{\odot}$ ($N(HII)$ $\lesssim$ 10$^2$). 
The difference in the number of HII regions is even more pronounced in bulge-dominated Sa-Sab systems\footnote{These morphological types
are lacking in the HI-rich sample. We do not expect that this selection effect can bias the previous conclusion since the calibration of the HI-deficiency parameter
has been done separately for each morphological class.}. The lower panel of Fig. \ref{scalingmass} clearly shows that the residual of this relation
($\Delta(y)$ ) is tightly correlated with the HI-deficiency parameter in particular in massive and intermediate mass systems ($M_{star}$ $\gtrsim$ 10$^8$ M$_{\odot}$), 
thus suggesting that for a given stellar mass the number of star forming regions decreases with the decrease of the atomic gas content.

   \begin{figure}
   \centering
   \includegraphics[width=0.43\textwidth]{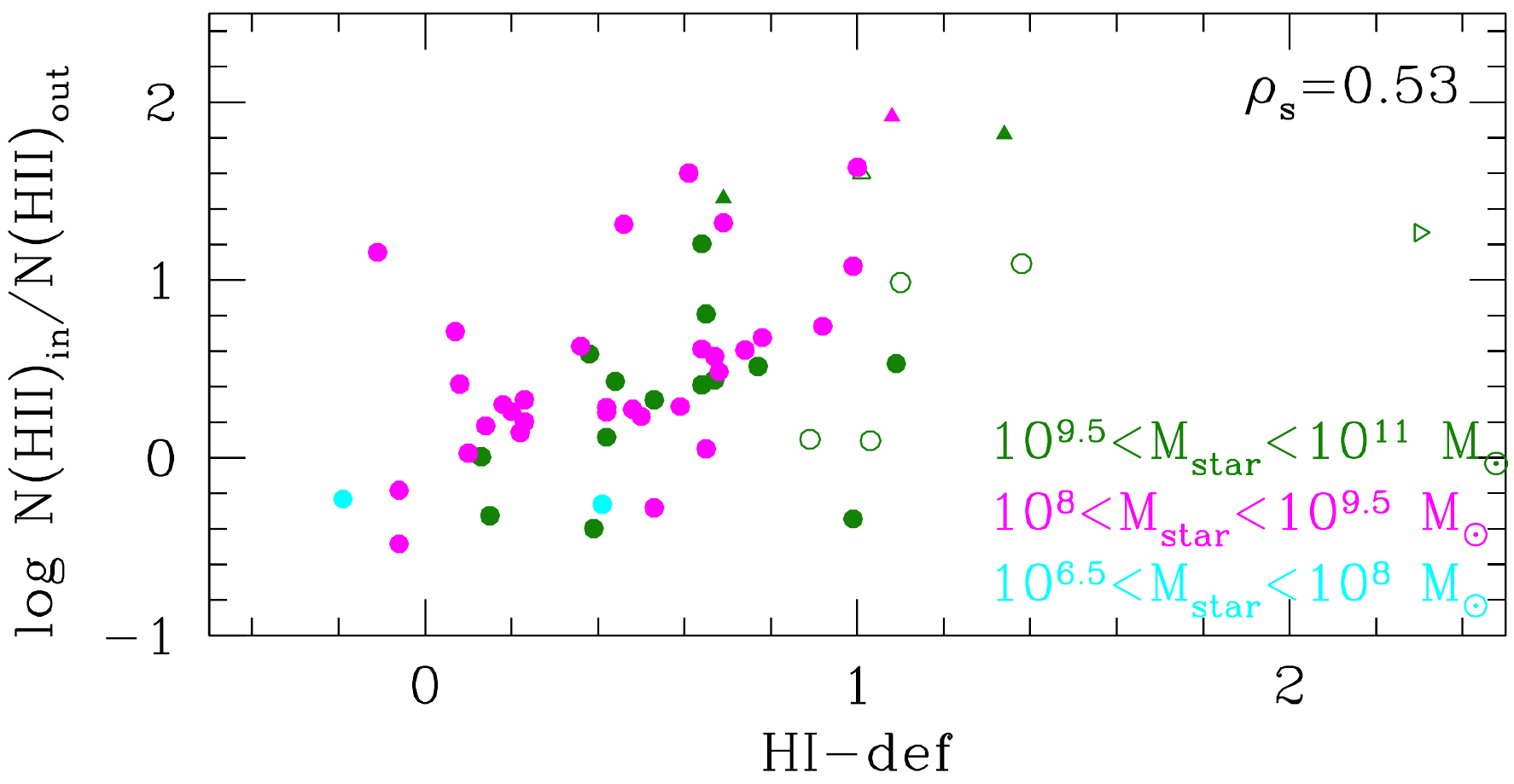}
   \caption{Relationship between the ratio of the number of HII regions of luminosity 
   $L(H\alpha)$ $>$ 10$^{37}$ erg s$^{-1}$ (for those galaxies with at least 20 HII regions)
   located within and outside the $r$-band effective radius and the HI-deficiency parameter. 
   Triangles are lower limits in the HI-deficiency parameter (when pointing to the right) or in the $N(HII)_{in}/N(HII)_{out}$ (when pointing to the top),
   these last indicating those objects where HII regions have been detected only within the $r$-band effective radius ($N(HII)_{out}$ = 0). Symbols are colour-coded as in Fig. \ref{scalingmass}.
   }
   \label{inout}%
   \end{figure}

Figure \ref{inout} shows the relationship between the ratio of the number of HII regions of luminosity $L(H\alpha)$ $>$ 10$^{37}$ erg s$^{-1}$ located 
within and outside the $r$-band effective radius but over the stellar disc of the galaxy, and the HI-deficiency parameter. To reduce the effects of the low number statistics, 
this plot is limited to those galaxies having at least 20 HII regions brighter than $L(H\alpha)$ $>$ 10$^{37}$ erg s$^{-1}$  (56 objects). 
%thus excluding low-mass systems ($M_{star}$ $\leq$ 10$^8$ M$_{\odot}$).
Again there is a statistically significant trend between the two variables (Spearman correlation coefficient $\rho_s$ = 0.60, corresponding 
to a $p$-value of 1.3 $\times$ 10$^{-5}$ for the alternative hypothesis of no correlation) 
indicating that in HI-deficient galaxies the number of HII regions is significantly reduced in the outer galaxy disc with respect to unperturbed systems.
The trend is even clearer if bulge-dominated galaxies (Sa-Sab), where the star formation activity might be reduced in the inner regions by the presence of the bulge, are removed.
Figure \ref{inout} being in logarithmic scale, in HI-deficient objects the number of HII regions in the outer disc drops by a factor of $\simeq$ 10 
with respect to that in the inner disc.  

   \begin{figure*}
   \centering
   \includegraphics[width=0.6\textwidth]{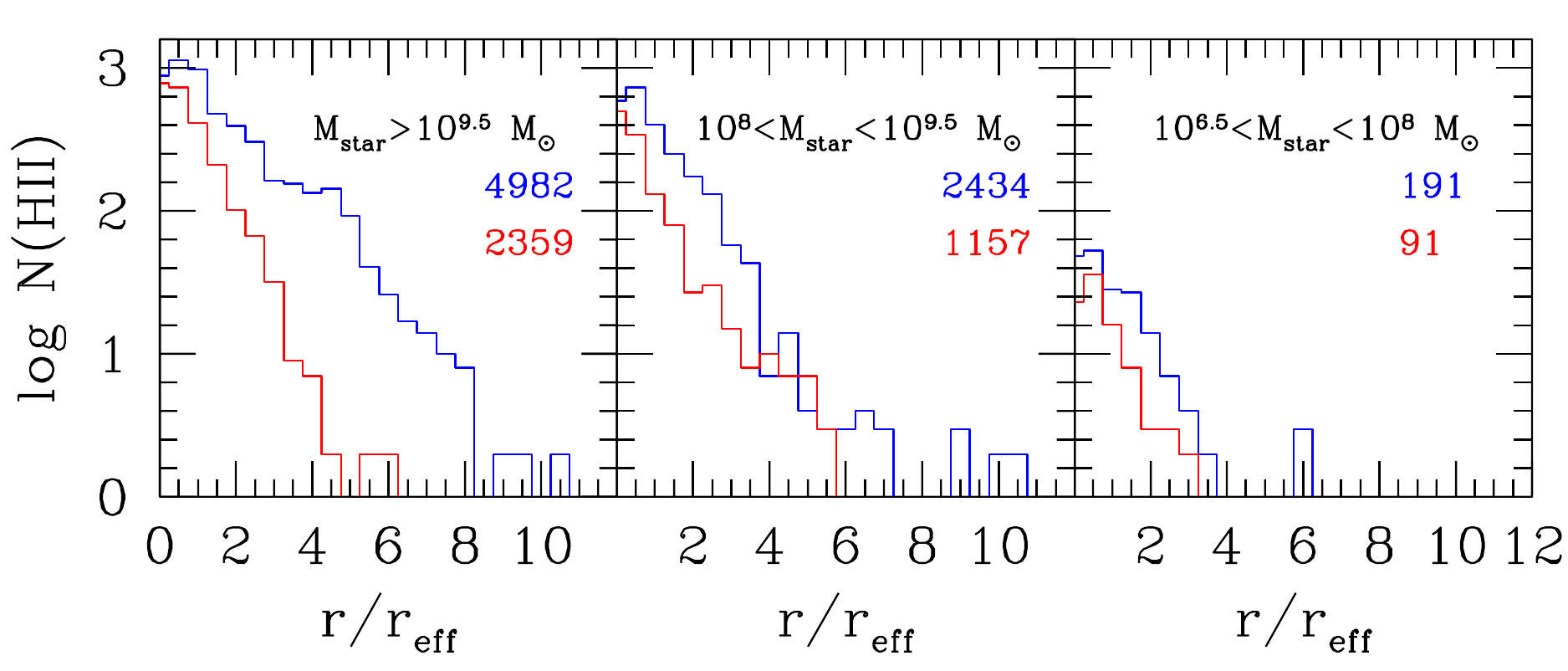}
   \caption{Radial distribution of the number of HII regions of luminosity $L(H\alpha)$ $>$ 10$^{37}$ erg s$^{-1}$ in units of effective radii for massive ($M_{star}$ $>$ 10$^{9.5}$ M$_{\odot}$ -
   left panel), intermediate (10$^{8}$ $<$ $M_{star}$ $\leq$ 10$^{9.5}$ M$_{\odot}$ - middle panel), and low mass galaxies (10$^{6.5}$ $<$ $M_{star}$ $\leq$ 10$^{8}$ M$_{\odot}$ - right panel).
   The blue and red values give the numbers of HII regions used to construct the histograms for HI-normal ($HI-def$ $\leq$ 0.5) and HI-deficient galaxies ($HI-def$ $>$ 0.5),
   respectively. 
   }
   \label{radiusdist}%
   \end{figure*}

Figure \ref{radiusdist} shows the radial distribution of the number of HII regions in units of effective radii for galaxies in different bins of stellar mass. Figure \ref{radiusdist} 
again indicates that the total number of HII regions is systematically lower in HI-stripped than in HI-normal galaxies, in particular in high-mass
systems ($M_{star}$ $>$ 10$^{9.5}$ M$_{\odot}$), and this despite the number of HI-deficient objects (20) is significantly higher than that of HI-normal galaxies (6).
It also shows that, in massive objects, the radial distribution of HII regions decreases more rapidly in HI-deficient than in HI-normal systems. 
A milder trend is also seen in the intermediate mass range, where several HII regions are present at galactocentric distances $r/r_{eff}$ $\gtrsim$ 6 only in HI-normal galaxies,
while nothing clear is seen in the low-mass range.

\section{Discussion}

The present set of data allows us to measure for the first time
ever the properties of HII regions located within the stellar disc of a statistically significant sample of cluster galaxies. \textit{The analysis of this unique set of data shows
that the number of HII regions is significantly reduced (more than a factor of ten) in gas-stripped galaxies, and that this occurs preferentially 
in the outer disc thus indicating an outside-in quenching of the star formation activity of cluster objects}. This decrease can hardly be
due to the growing importance of the bulge or the presence of bars, which might affect only the star formation activity in the inner regions
(Kennicutt et al. 1994, Gavazzi et al. 2015, Consolandi et al. 2017). If overall this result is expected given the physical link between gas content and star formation, 
it is worth noticing that a tight relation between star formation and gas column density (Schmidt law) is observed mainly when the molecular gas phase is considered (e.g. Bigiel et al. 2008).
Furthermore, the HI-deficiency parameter is tightly related to the size of the HI and H$_2$ discs (Vollmer et al. 2001, Boselli et al. 2014a). This
global indicator is thus very sensitive to the presence of gas in the outer disc where star formation does not necessarily occur. 
In other words, the column density of the molecular gas in the inner regions, where star formation takes place, must be somehow related to the total amount of atomic gas
available on the HI disc (e.g. Boselli et al. 2001), which in field galaxies generally extends up to $\sim$ 1.8 times the stellar disc (Cayatte et al. 1994).
  
This new result is consistent with what is observed using integrated quantities (e.g. HI-deficient galaxies located below the main sequence relation,
Boselli et al. 2015; HI-deficient galaxies mainly located in the green valley, Boselli et al. 2014a,b; decrease of the specific star formation rate in 
HI-deficient galaxies, Gavazzi et al. 2013) and with the morphological analysis of the star forming discs of Virgo cluster galaxies (presence of truncated H$\alpha$ and UV discs in HI-deficient galaxies,
Koopmann \& Kenney 1998, Koopmann et al. 2001, Fossati et al. 2012, Boselli et al. 2015, Cortese et al. 2012b; outside-in quenching of the 
star formation activity, Boselli et al. 2006, Fossati et al. 2018). Compared to these previous studies, the present analysis clearly demonstrates that the reduced star 
formation activity is related to a reduced number of HII regions in gas-stripped objects, and thus that the perturbing mechanism acts down to scales $\lesssim$ 100 pc,
which is the typical scale of giant molecular clouds.  
The reduced number of HII regions in HI-deficient galaxies primarily occurring in their outer disc is a further indication that 
the main quenching process is ram pressure stripping.  
This mechanism removes the gas necessary to feed star formation outside-in, producing
truncated gas discs. Truncated discs are indeed observed in the different components of the ISM that are well resolved at all frequencies in many Virgo 
cluster galaxies (HI: Cayatte et al. 1990; Chung et al. 2009; H$_2$: Fumagalli et al. 2009, Boselli et al. 2014a; dust: Cortese et al. 2010). 
The lack of gas in the outer regions thus prevents the formation of new stars. %, or in other words, the Schmidt law is independent on the environment
%and the decrease in the star formation activity is just due to a decrease of the gas column density.
It also rules out milder mechanisms such as starvation or strangulation, where the cessation of gas infall would 
induce a similar quenching of the star formation activity at all galactocentric distances (Boselli et al. 2006). 
%The large scatter in the relation shown in Fig. \ref{inout}, however, suggests that other mechanisms can also participate or contribute to the 
%stripping process, as indeed observed in several well studied objects (NGC 4438, Boselli et al. 2005, Vollmer et al. 2009; NGC 4654, Vollmer et al. 2003; NGC 4424, 
%Cortes et al. 2006, Sorgho et al. 2017, Boselli et al. 2018c).

The different behaviour in Fig. \ref{radiusdist} of dwarfs ($M_{star}$ $\lesssim$ 10$^8$ M$_{\odot}$) vs. massive systems
might be ascribed to a joint effect of poor number statistics (30/32 galaxies in this mass range have less than 20 HII regions) and to 
an inaccurate calibration of the HI-deficiency parameter in the undersampled dwarf galaxy population with HI data in the field (Boselli \& Gavazzi 2009)
and where the very irregular nature of the galaxies define highly scattered scaling relations.
It could also result, however, from the fact that the typical time-scale for gas stripping and quenching of the star formation activity ($\simeq$ 100 Myr), 
is significantly longer than the life of individual HII regions ($\lesssim$ 10 Myr) only in massive systems, while it is comparable in dwarfs where the gas loosely bound
to the shallow potential well of the galaxy is rapidly stripped transforming star-forming systems into quiescent dwarf ellipticals (Boselli et al. 2008).
%Figures \ref{scalingmass} and \ref{inout} also suggest that the quenching of the star formation activity of the gas-stripped galaxies is due to a decrease 
%of the number of HII regions rather than to a decrease of the number of newly formed stars inside each single HII region.
%This would imply that the quenching mechanism does not affect the star formation process within individual HII regions, which behave as similar objects in unperturbed systems,
%but just reduce their total number.

\section{Conclusions}

The exceptional quality of the VESTIGE survey in terms of angular resolution and sensitivity 
allow us to attempt the first statistical study on the effects of the environment 
on the star formation process on cluster galaxies down to the scale of individual HII regions. Using the \textsc{HIIphot} 
code we detect 11302 HII regions of luminosity $L(H\alpha)$ $>$ 10$^{37}$ erg s$^{-1}$ on a sample 
of 114 late-type galaxies observed so far in the Virgo cluster. We derive the typical scaling relation between the number of 
HII regions and the stellar mass of the sample galaxies, and we show that those objects stripped of their gas during the interaction
with the hostile cluster environment have on average 
a lower number of HII regions than gas rich systems of similar stellar mass. We also show that in HI-stripped galaxies the number of HII
regions is mainly reduced with respect to that of HI-rich systems in the outer discs. These results, consistently with previous results
based on the multifrequency analysis of large samples of galaxies or dedicated studies of representative objects,
indicate that in a cluster as massive as Virgo  
the dominant perturbing mechanism responsible for the quenching of the star formation activity is ram pressure stripping.
This promising result once again stresses the uniqueness of complete deep surveys of nearby targets for the study of the different 
mechanisms shaping galaxy evolution.

\begin{acknowledgements}

We thank D. Thilker for providing us with the \textsc{HIIphot} data reduction pipeline and
M. Balogh, F. Combes, J. Roediger, J. Taylor, S. Tonnesen, and the referee for constructive comments and suggestions. 
We are grateful to the CFHT team who assisted us in the observations: 
T. Burdullis, D. Devost, B. Mahoney, N. Manset, A. Petric, S. Prunet, K. Withington.
We acknowledge financial support from "Programme National de Cosmologie and Galaxies" (PNCG) 
and from "Projet International de Coop\'eration Scientifique" (PICS) with Canada.
MF has received funding from the European Research Council (ERC) (grant agreement No 757535),
AL from the French Centre National d'Etudes Spatiales (CNES).

%This project has received funding from the European Research Council (ERC) under the European Union's Horizon 2020 research and innovation programme (grant agreement No 757535 and
%grant agreement No 679627, project name FORNAX).
%MB acknowledges the FONDECYT regular grant 1170618.
%MS acknowledges support from the NSF grant 1714764.
%MB was supported by MINEDUC-UA projects, code ANT 1655 and ANT 1656.
%AL work was supported by the Sino-French LIA-Origin joint program.
%MS acknowledges support from the NSF grant 1714764 and the Chandra Award GO6-17111X.
%MF acknowledges support by the science and technology facilities council [grant number ST/P000541/1].
%EWP acknowledges support from the National Natural Science Foundation of China through Grant No. 11573002.
%KS acknowledges support from the Natural Sciences and Engineering Research Council of Canada (NSERC).

\end{acknowledgements}


\begin{thebibliography}{}

%\bibitem[Abraham et al.(2003)]{2003ApJ...588..218A} Abraham, R.~G., van den Bergh, S., \& Nair, P.\ 2003, \apj, 588, 218 
%\bibitem[Abramson \& Kenney(2014)]{2014AJ....147...63A} Abramson, A., \& Kenney, J.~D.~P.\ 2014, \aj, 147, 63 
%\bibitem[Abramson et al.(2011)]{2011AJ....141..164A} Abramson, A., Kenney, J.~D.~P., Crowl, H.~H., et al.\ 2011, \aj, 141, 164 
%\bibitem[Abramson et al.(2016)]{2016AJ....152...32A} Abramson, A., Kenney, J., Crowl, H., \& Tal, T.\ 2016, \aj, 152, 32 
%\bibitem[Adams et al.(2013)]{2013ApJ...768...77A} Adams, E.~A.~K., Giovanelli, R., \& Haynes, M.~P.\ 2013, \apj, 768, 77 
%\bibitem[Aguerri et al.(2005)]{2005AJ....129.2585A} Aguerri, J.~A.~L., Gerhard, O.~E., Arnaboldi, M., et al.\ 2005, \aj, 129, 2585 
%\bibitem[Albareti et al. (2017)]{} Albareti, F., Allende Prieto, C., Almeida A., et al.\ 2017, arXiv:1608.02013  
%\bibitem[Ambrocio-Cruz et al.(2016)]{2016MNRAS.457.2048A} Ambrocio-Cruz, P., Le Coarer, E., Rosado, M., et al.\ 2016, \mnras, 457, 2048 
%\bibitem[Anderson \& Sunayaev (2018)]{}Anderson \& Sunayaev 2018, https://arxiv.org/abs/1712.07535
%\bibitem[Arnaboldi et al.(1996)]{1996ApJ...472..145A} Arnaboldi, M., Freeman, K.~C., Mendez, R.~H., et al.\ 1996, \apj, 472, 145 
%\bibitem[Arnaboldi et al.(2002)]{2002AJ....123..760A} Arnaboldi, M., Aguerri, J.~A.~L., Napolitano, N.~R., et al.\ 2002, \aj, 123, 760 
%\bibitem[Arnaboldi et al.(2003)]{2003AJ....125..514A} Arnaboldi, M., Freeman, K.~C., Okamura, S., et al.\ 2003, \aj, 125, 514 
%\bibitem[Arnaboldi et al.(2004)]{2004ApJ...614L..33A} Arnaboldi, M., Gerhard, O., Aguerri, J.~A.~L., et al.\ 2004, \apjl, 614, L33 
%\bibitem[Arnold et al.(2014)]{2014ApJ...791...80A} Arnold, J.~A., Romanowsky, A.~J., Brodie, J.~P., et al.\ 2014, \apj, 791, 80 
%\bibitem[Arp(1967)]{1967ApL.....1....1A} Arp, H.~C.\ 1967, \aplett, 1, 1 
%\bibitem[Arrigoni Battaia et al.(2012)]{2012A&A...543A.112A} Arrigoni Battaia, F., Gavazzi, G., Fumagalli, M., et al.\ 2012, \aap, 543, A112 
%\bibitem[Auld et al.(2013)]{2013MNRAS.428.1880A} Auld, R., Bianchi, S., Smith, M.~W.~L., et al.\ 2013, \mnras, 428, 1880 
\bibitem[Azimlu et al.(2011)]{2011AJ....142..139A} Azimlu, M., Marciniak, R., \& Barmby, P.\ 2011, \aj, 142, 139
%\bibitem[Baade \& Minkowski(1954)]{1954ApJ...119..206B} Baade, W., \& Minkowski, R.\ 1954, \apj, 119, 206 
%\bibitem[Bacon et al.(2017)]{2017arXiv171003002B} Bacon, R., Conseil, S., Mary, D., et al.\ 2017, arXiv:1710.03002 
%\bibitem[Baes et al.(2010)]{2010A&A...518L..53B} Baes, M., Clemens, M., Xilouris, E.~M., et al.\ 2010, \aap, 518, L53 
%\bibitem[Baes et al.(2014)]{2014A&A...562A.106B} Baes, M., Herranz, D., Bianchi, S., et al.\ 2014, \aap, 562, A106 
%\bibitem[Baldi et al.(2009)]{2009ApJ...707.1034B} Baldi, A., Forman, W., Jones, C., et al.\ 2009, \apj, 707, 1034 
%\bibitem[Baldry et al.(2006)]{2006MNRAS.373..469B} Baldry, I.~K., Balogh, M.~L., Bower, R.~G., et al.\ 2006, \mnras, 373, 469 
%\bibitem[Baldwin et al.(1981)]{1981PASP...93....5B} Baldwin, J.~A., Phillips, M.~M., \& Terlevich, R.\ 1981, \pasp, 93, 5 
%\bibitem[Balmaverde et al.(2018)]{2018A&A...612A..19B} Balmaverde, B., Capetti, A., Marconi, A., \& Venturi, G.\ 2018, \aap, 612, A19 
%\bibitem[Balogh et al.(2000)]{2000ApJ...540..113B} Balogh, M.~L., Navarro, J.~F., \& Morris, S.~L.\ 2000, \apj, 540, 113 
%\bibitem[Bamford et al.(2009)]{2009MNRAS.393.1324B} Bamford, S.~P., Nichol, R.~C., Baldry, I.~K., et al.\ 2009, \mnras, 393, 1324 
%\bibitem[Barnes et al.(2017)]{2017MNRAS.471.1088B} Barnes, D.~J., Kay, S.~T., Bah{\'e}, Y.~M., et al.\ 2017, \mnras, 471, 1088 
%\bibitem[Barton et al.(2000)]{2000ApJ...530..660B} Barton, E.~J., Geller, M.~J., \& Kenyon, S.~J.\ 2000, \apj, 530, 660 
%\bibitem[Baum et al.(1988)]{1988ApJS...68..643B} Baum, S.~A., Heckman, T.~M., Bridle, A., van Breugel, W.~J.~M., \& Miley, G.~K.\ 1988, \apjs, 68, 643 
%\bibitem[Becker et al.(1995)]{1995ApJ...450..559B} Becker, R.~H., White, R.~L., \& Helfand, D.~J.\ 1995, \apj, 450, 559 
%\bibitem[Beckman et al.(2000)]{2000AJ....119.2728B} Beckman, J.~E., Rozas, M., Zurita, A., Watson, R.~A., \& Knapen, J.~H.\ 2000, \aj, 119, 2728 
%\bibitem[Bekki(2009)]{2009MNRAS.399.2221B} Bekki, K.\ 2009, \mnras, 399, 2221 
%\bibitem[Bekki(2014)]{2014MNRAS.438..444B} Bekki, K.\ 2014, \mnras, 438, 444 
%\bibitem[Belfiore et al.(2016)]{2016MNRAS.461.3111B} Belfiore, F., Maiolino, R., Maraston, C., et al.\ 2016, \mnras, 461, 3111 
%\bibitem[Bellazzini et al.(2015)]{2015ApJ...800L..15B} Bellazzini, M., Magrini, L., Mucciarelli, A., et al.\ 2015, \apjl, 800, L15 
%\bibitem[Bellhouse et al.(2017)]{2017ApJ...844...49B} Bellhouse, C., Jaff{\'e}, Y.~L., Hau, G.~K.~T., et al.\ 2017, \apj, 844, 49 
%\bibitem[Bellini et al.(2015)]{2015ApJ...805..178B} Bellini, A., Renzini, A., Anderson, J., et al.\ 2015, \apj, 805, 178 
%\bibitem[Bendo et al.(2012)]{2012MNRAS.423..197B} Bendo, G.~J., Galliano, F., \& Madden, S.~C.\ 2012, \mnras, 423, 197 
%\bibitem[Bertin \& Arnouts(1996)]{1996A&AS..117..393B} Bertin, E., \& Arnouts, S.\ 1996, \aaps, 117, 393 
%\bibitem[Bianchi et al.(2017)]{2017A&A...597A.130B} Bianchi, S., Giovanardi, C., Smith, M.~W.~L., et al.\ 2017, \aap, 597, A130 
\bibitem[Bigiel et al.(2008)]{2008AJ....136.2846B} Bigiel, F., Leroy, A., Walter, F., et al.\ 2008, \aj, 136, 2846
%\bibitem[Biller et al.(2004)]{2004ApJ...613..238B} Biller, B.~A., Jones, C., Forman, W.~R., Kraft, R., \& Ensslin, T.\ 2004, \apj, 613, 238 
\bibitem[Binggeli et al.(1985)]{1985AJ.....90.1681B} Binggeli, B., Sandage, A., \& Tammann, G.~A.\ 1985, \aj, 90, 1681 
%\bibitem[Binggeli et al.(1987)]{1987AJ.....94..251B} Binggeli, B., Tammann, G.~A., \& Sandage, A.\ 1987, \aj, 94, 251 
%\bibitem[Binney et al.(2009)]{2009MNRAS.397.1804B} Binney, J., Nipoti, C., \& Fraternali, F.\ 2009, \mnras, 397, 1804 
%\bibitem[Biviano et al.(2006)]{2006A&A...456...23B} Biviano, A., Murante, G., Borgani, S., et al.\ 2006, \aap, 456, 23 
\bibitem[Blakeslee et al.(2009)]{2009ApJ...694..556B} Blakeslee, J.~P., Jord{\'a}n, A., Mei, S., et al.\ 2009, \apj, 694, 556 
%\bibitem[Bland-Hawthorn \& Maloney(1999)]{1999ApJ...510L..33B} Bland-Hawthorn, J., \& Maloney, P.~R.\ 1999, \apjl, 510, L33 
%\bibitem[Blanton et al.(2005)]{2005ApJ...629..143B} Blanton, M.~R., Eisenstein, D., Hogg, D.~W., Schlegel, D.~J., \& Brinkmann, J.\ 2005, \apj, 629, 143 
%\bibitem[B{\"o}hringer et al.(1994)]{1994Natur.368..828B} B{\"o}hringer, H., Briel, U.~G., Schwarz, R.~A., et al.\ 1994, \nat, 368, 828 
%\bibitem[Boissier et al.(2001)]{2001MNRAS.321..733B} Boissier, S., Boselli, A., Prantzos, N., \& Gavazzi, G.\ 2001, \mnras, 321, 733 
%\bibitem[Boissier et al.(2003)]{2003MNRAS.346.1215B} Boissier, S., Prantzos, N., Boselli, A., \& Gavazzi, G.\ 2003a, \mnras, 346, 1215 
%\bibitem[Boissier et al.(2003)]{2003MNRAS.343..653B} Boissier, S., Monnier Ragaigne, D., Prantzos, N., et al.\ 2003b, \mnras, 343, 653 
%\bibitem[Boissier et al.(2007)]{2007ApJS..173..524B} Boissier, S., Gil de Paz, A., Boselli, A., et al.\ 2007, \apjs, 173, 524 
%\bibitem[Boissier et al.(2008)]{2008ApJ...681..244B} Boissier, S., Gil de Paz, A., Boselli, A., et al.\ 2008, \apj, 681, 244-257 
%\bibitem[Boissier et al.(2012)]{2012A&A...545A.142B} Boissier, S., Boselli, A., Duc, P.-A., et al.\ 2012, \aap, 545, A142 
%\bibitem[Boissier et al.(2015)]{2015A&A...579A..29B} Boissier, S., Boselli, A., Voyer, E., et al.\ 2015, \aap, 579, A29 
%\bibitem[Bogd{\'a}n et al.(2012)]{2012ApJ...755...25B} Bogd{\'a}n, {\'A}., Forman, W.~R., Kraft, R.~P., et al.\ 2012, \apj, 755, 25 
%\bibitem[Boquien et al.(2012)]{2012A&A...539A.145B} Boquien, M., Buat, V., Boselli, A., et al.\ 2012, \aap, 539, A145 
%\bibitem[Boquien et al.(2014)]{2014A&A...571A..72B} Boquien, M., Buat, V., \& Perret, V.\ 2014, \aap, 571, A72 
%\bibitem[Boquien et al.(2015)]{2015A&A...578A...8B} Boquien, M., Calzetti, D., Aalto, S., et al.\ 2015, \aap, 578, A8 
%\bibitem[Boquien et al.(2016)]{2016A&A...591A...6B} Boquien, M., Kennicutt, R., Calzetti, D., et al.\ 2016, \aap, 591, A6 
%\bibitem[Boroson et al.(1983)]{1983AJ.....88.1707B} Boroson, T.~A., Thompson, I.~B., \& Shectman, S.~A.\ 1983, \aj, 88, 1707 
%\bibitem[Boselli(2011)]{2011pvg..book.....B} Boselli, A.\ 2011, A Panchromatic View of Galaxies, by Alessandro Boselli.~- Practical Approach Book - ISBN-10: 3-527-40991-2.~ISBN-13: 978-3-527-40991-4 - Wiley-VCH, Berlin 2011.~XVI, 324pp, Hardcover,  
%\bibitem[Boselli \& Gavazzi(2002)]{2002A&A...386..124B} Boselli, A., \& Gavazzi, G.\ 2002, \aap, 386, 124 
\bibitem{2006PASP..118..517B} Boselli, A., \& Gavazzi, G.\ 2006, \pasp, 118, 517 
\bibitem[Boselli, \& Gavazzi(2009)]{2009A&A...508..201B} Boselli, A., \& Gavazzi, G.\ 2009, \aap, 508, 201
\bibitem[Boselli \& Gavazzi(2014)]{2014A&ARv..22...74B} Boselli, A., \& Gavazzi, G.\ 2014, \aapr, 22, 74 
%\bibitem[Boselli et al.(1995)]{1995A&AS..110..521B} Boselli, A., Casoli, F., \& Lequeux, J.\ 1995, \aaps, 110, 521 
\bibitem[Boselli et al.(2001)]{2001AJ....121..753B} Boselli, A., Gavazzi, G., Donas, J., \& Scodeggio, M.\ 2001, \aj, 121, 753 
%\bibitem[Boselli et al.(2002)]{2002A&A...386..134B} Boselli, A., Iglesias-P{\'a}ramo, J., V{\'{\i}}lchez, J.~M., \& Gavazzi, G.\ 2002b, \aap, 386, 134 
%\bibitem[Boselli et al.(2002)]{2002A&A...384...33B} Boselli, A., Lequeux, J., \& Gavazzi, G.\ 2002a, \aap, 384, 33 
%\bibitem[Boselli et al.(2003)]{2003A&A...402...37B} Boselli, A., Gavazzi, G., \& Sanvito, G.\ 2003a, \aap, 402, 37 
%\bibitem[Boselli et al.(2003)]{2003A&A...406..867B} Boselli, A., Sauvage, M., Lequeux, J., Donati, A., \& Gavazzi, G.\ 2003, \aap, 406, 867 
%\bibitem[Boselli et al.(2005)]{2005ApJ...629L..29B} Boselli, A., Cortese, L., Deharveng, J.~M., et al.\ 2005, \apjl, 629, L29 
%\bibitem[Boselli et al.(2005)]{2005ApJ...623L..13B} Boselli, A., Boissier, S., Cortese, L., et al.\ 2005, \apjl, 623, L13 
\bibitem{2006ApJ...651..811B} Boselli, A., Boissier, S., Cortese, L., et al.\ 2006, \apj, 651, 811 
\bibitem{2008ApJ...674..742B} Boselli, A., Boissier, S., Cortese, L., \& Gavazzi, G.\ 2008, \apj, 674, 742 
%\bibitem{2008A&A...489.1015B} Boselli, A., Boissier, S., Cortese, L., \& Gavazzi, G.\ 2008b, \aap, 489, 1015 
\bibitem[Boselli et al.(2009)]{2009ApJ...706.1527B} Boselli, A., Boissier, S., Cortese, L., et al.\ 2009, \apj, 706, 1527 
%\bibitem[Boselli et al.(2010)]{2010PASP..122..261B} Boselli, A., Eales, S., Cortese, L., et al.\ 2010a, \pasp, 122, 261 
%\bibitem[Boselli et al.(2010)]{2010A&A...518L..61B} Boselli, A., Ciesla, L., Buat, V., et al.\ 2010, \aap, 518, L61 
%\bibitem{2011A&A...528A.107B} Boselli, A., Boissier, S., Heinis, S., et al.\ 2011, \aap, 528, A107 
%\bibitem[Boselli et al.(2013)]{2013A&A...550A.114B} Boselli, A., Hughes, T.~M., Cortese, L., Gavazzi, G., \& Buat, V.\ 2013, \aap, 550, A114 
%\bibitem[Boselli et al.(2014)]{2014A&A...564A..65B} Boselli, A., Cortese, L., \& Boquien, M.\ 2014b, \aap, 564, A65 
\bibitem[Boselli et al.(2014)]{2014A&A...564A..67B} Boselli, A., Cortese, L., Boquien, M., et al.\ 2014a, \aap, 564, A67 
\bibitem[Boselli et al.(2014)]{2014A&A...570A..69B} Boselli, A., Voyer, E., Boissier, S., et al.\ 2014b, \aap, 570, AA69 
\bibitem[Boselli et al.(2015)]{2015A&A...579A.102B} Boselli, A., Fossati, M., Gavazzi, G., et al.\ 2015, \aap, 579, A102 
\bibitem[Boselli et al.(2016)]{2016A&A...596A..11B} Boselli, A., Roehlly, Y., Fossati, M., et al.\ 2016, \aap, 596, A11 
%\bibitem[Boselli et al.(2016)]{2016A&A...587A..68B} Boselli, A., Cuillandre, J.~C., Fossati, M., et al.\ 2016b, \aap, 587, A68 
%\bibitem[Boselli et al.(2016)]{2016A&A...585A...2B} Boselli, A., Boissier, S., Voyer, E., et al.\ 2016c, \aap, 585, A2 
%\bibitem[Boselli et al.(2017)]{} Boselli A., Fossati M., Cuillandre J.C., et al., submitted to A\&A (paper IV)
\bibitem[Boselli et al.(2018)]{2018A&A...614A..56B} Boselli, A., Fossati, M., Ferrarese, L., et al.\ 2018, \aap, 614, A56 
%\bibitem[Boselli et al.(2018)]{2018A&A...615A.114B} Boselli, A., Fossati, M., Cuillandre, J.~C., et al.\ 2018b, \aap, 615, A114 
%\bibitem[Boselli et al.(2018)]{2018A&A...620A.164B} Boselli, A., Fossati, M., Consolandi, G., et al.\ 2018c, \aap, 620, A164
\bibitem[Boselli et al.(2019)]{2019A&A...631A.114B} Boselli, A., Epinat, B., Contini, T., et al.\ 2019a, \aap, 631, A114
\bibitem[Boselli et al.(2019)]{2019A&A...623A..52B} Boselli, A., Fossati, M., Longobardi, A., et al.\ 2019b, \aap, 623, A52
%\bibitem[Boulade et al.(2003)]{2003SPIE.4841...72B} Boulade, O., Charlot, X., Abbon, P., et al.\ 2003, \procspie, 4841, 72 
%\bibitem[Boulanger et al.(1996)]{1996A&A...312..256B} Boulanger, F., Abergel, A., Bernard, J.-P., et al.\ 1996, \aap, 312, 256 
%\bibitem[Bovy(2015)]{2015ApJS..216...29B} Bovy, J.\ 2015, \apjs, 216, 29 
\bibitem[Bradley et al.(2006)]{2006A&A...459L..13B} Bradley, T.~R., Knapen, J.~H., Beckman, J.~E., \& Folkes, S.~L.\ 2006, \aap, 459, L13 
%\bibitem[Braine \& Wiklind(1993)]{1993A&A...267L..47B} Braine, J., \& Wiklind, T.\ 1993, \aap, 267, L47 
%\bibitem[Bryan et al.(2014)]{2014ApJS..211...19B} Bryan, G.~L., Norman, M.~L., O'Shea, B.~W., et al.\ 2014, \apjs, 211, 19 
%\bibitem[Buckalew \& Kobulnicky(2006)]{2006AJ....132.1061B} Buckalew, B.~A., \& Kobulnicky, H.~A.\ 2006, \aj, 132, 1061 
%\bibitem[Bullock(2010)]{2010arXiv1009.4505B} Bullock, J.~S.\ 2010, arXiv:1009.4505 
%\bibitem[Burkhart \& Loeb(2016)]{2016ApJ...824L...7B} Burkhart, B., \& Loeb, A.\ 2016, \apjl, 824, L7 
%\bibitem[Buzzoni et al.(2006)]{2006MNRAS.368..877B} Buzzoni, A., Arnaboldi, M., \& Corradi, R.~L.~M.\ 2006, \mnras, 368, 877 
\bibitem[Byrd \& Valtonen(1990)]{1990ApJ...350...89B} Byrd, G., \& Valtonen, M.\ 1990, \apj, 350, 89 
%\bibitem[Calzetti et al.(2007)]{2007ApJ...666..870C} Calzetti, D., Kennicutt, R.~C., Engelbracht, C.~W., et al.\ 2007, \apj, 666, 870 
%\bibitem[Calzetti et al.(2010)]{2010ApJ...714.1256C} Calzetti, D., Wu, S.-Y., Hong, S., et al.\ 2010, \apj, 714, 1256 
%\bibitem[Canning et al.(2014)]{2014MNRAS.444..336C} Canning, R.~E.~A., Ryon, J.~E., Gallagher, J.~S., et al.\ 2014, \mnras, 444, 336 
%\bibitem[Cannon et al.(2015)]{2015AJ....149...72C} Cannon, J.~M., Martinkus, C.~P., Leisman, L., et al.\ 2015, \aj, 149, 72 
%\bibitem[Cappellari \& Emsellem(2004)]{2004PASP..116..138C} Cappellari, M., \& Emsellem, E.\ 2004, \pasp, 116, 138 
%\bibitem[Cappellari et al.(2011)]{2011MNRAS.413..813C} Cappellari, M., Emsellem, E., Krajnovi{\'c}, D., et al.\ 2011, \mnras, 413, 813 
%\bibitem[Castro-Rodrigu{\'e}z et al.(2009)]{2009A&A...507..621C} Castro-Rodrigu{\'e}z, N., Arnaboldi, M., Aguerri, J.~A.~L., et al.\ 2009, \aap, 507, 621 
\bibitem[Cayatte et al.(1990)]{1990AJ....100..604C} Cayatte, V., van Gorkom, J.~H., Balkowski, C., \& Kotanyi, C.\ 1990, \aj, 100, 604 
\bibitem[Cayatte et al.(1994)]{1994AJ....107.1003C} Cayatte, V., Kotanyi, C., Balkowski, C., et al.\ 1994, \aj, 107, 1003
%\bibitem[Cen(2014)]{2014ApJ...781...38C} Cen, R.\ 2014, \apj, 781, 38 
%\bibitem[Chamaraux et al.(1980)]{1980A&A....83...38C} Chamaraux, P., Balkowski, C., \& Gerard, E.\ 1980, \aap, 83, 38 
%\bibitem[Chemin et al.(2006)]{2006MNRAS.366..812C} Chemin, L., Balkowski, C., Cayatte, V., et al.\ 2006, \mnras, 366, 812 
%\bibitem[Chung et al.(2007)]{2007ApJ...659L.115C} Chung, A., van Gorkom, J.~H., Kenney, J.~D.~P., \& Vollmer, B.\ 2007, \apjl, 659, L115 
\bibitem[Chung et al.(2009)]{2009AJ....138.1741C} Chung, A., van Gorkom, J.~H., Kenney, J.~D.~P., Crowl, H., \& Vollmer, B.\ 2009a, \aj, 138, 1741 
%\bibitem[Chung et al.(2009)]{2009ApJS..184..199C} Chung, E.~J., Rhee, M.-H., Kim, H., et al.\ 2009b, \apjs, 184, 199 
%\bibitem[Churazov et al.(2001)]{2001ApJ...554..261C} Churazov, E., Br{\"u}ggen, M., Kaiser, C.~R., B{\"o}hringer, H., \& Forman, W.\ 2001, \apj, 554, 261 
%\bibitem[Churazov et al.(2008)]{2008MNRAS.388.1062C} Churazov, E., Forman, W., Vikhlinin, A., et al.\ 2008, \mnras, 388, 1062 
%\bibitem[Ciambur(2015)]{2015ApJ...810..120C} Ciambur, B.~C.\ 2015, \apj, 810, 120 
%\bibitem[Ciardullo(2010)]{2010PASA...27..149C} Ciardullo, R.\ 2010, \pasa, 27, 149 
%\bibitem[Ciardullo et al.(1998)]{1998ApJ...492...62C} Ciardullo, R., Jacoby, G.~H., Feldmeier, J.~J., \& Bartlett, R.~E.\ 1998, \apj, 492, 62 
%\bibitem[Ciardullo et al.(2013)]{2013ApJ...769...83C} Ciardullo, R., Gronwall, C., Adams, J.~J., et al.\ 2013, \apj, 769, 83 
%\bibitem[Ciesla et al.(2012)]{2012A&A...543A.161C} Ciesla, L., Boselli, A., Smith, M.~W.~L., et al.\ 2012, \aap, 543, A161 
%\bibitem[Ciesla et al.(2014)]{2014A&A...565A.128C} Ciesla, L., Boquien, M., Boselli, A., et al.\ 2014, \aap, 565, A128 
%\bibitem[Combes et al.(2007)]{2007MNRAS.377.1795C} Combes, F., Young, L.~M., \& Bureau, M.\ 2007, \mnras, 377, 1795 
%\bibitem[Condon et al.(1998)]{1998AJ....115.1693C} Condon, J.~J., Cotton, W.~D., Greisen, E.~W., et al.\ 1998, \aj, 115, 1693 
%\bibitem[Conselice et al.(2001)]{2001AJ....122.2281C} Conselice, C.~J., Gallagher, J.~S., III, \& Wyse, R.~F.~G.\ 2001, \aj, 122, 2281 
%\bibitem[Conselice(2003)]{2003ApJS..147....1C} Conselice, C.~J.\ 2003, \apjs, 147, 1 
%\bibitem[Consolandi(2016)]{2016A&A...595A..67C} Consolandi, G.\ 2016, \aap, 595, A67 
%\bibitem[Consolandi et al.(2016)]{2016A&A...591A..38C} Consolandi, G., Gavazzi, G., Fumagalli, M., Dotti, M., \& Fossati, M.\ 2016, \aap, 591, A38 
\bibitem[Consolandi et al.(2017)]{2017A&A...598A.114C} Consolandi, G., Dotti, M., Boselli, A., Gavazzi, G., \& Gargiulo, F.\ 2017, \aap, 598, A114 
%\bibitem[Consolandi et al.(2017)]{2017A&A...606A..83C} Consolandi, G., Gavazzi, G., Fossati, M., et al.\ 2017, \aap, 606, A83 
%\bibitem[Contini et al.(2014)]{2014MNRAS.437.3787C} Contini, E., De Lucia, G., Villalobos, {\'A}., \& Borgani, S.\ 2014, \mnras, 437, 3787 
%\bibitem[Cort{\'e}s et al.(2006)]{2006AJ....131..747C} Cort{\'e}s, J.~R., Kenney, J.~D.~P., \& Hardy, E.\ 2006, \aj, 131, 747
%\bibitem[Cort{\'e}s et al.(2015)]{2015ApJS..216....9C} Cort{\'e}s, J.~R., Kenney, J.~D.~P., \& Hardy, E.\ 2015, \apjs, 216, 9 
%\bibitem[Cortese et al.(2006)]{2006A&A...453..847C} Cortese, L., Gavazzi, G., Boselli, A., et al.\ 2006, \aap, 453, 847 
\bibitem[Cortese et al.(2010)]{2010A&A...518L..49C} Cortese, L., Davies, J.~I., Pohlen, M., et al.\ 2010a, \aap, 518, L49 
%\bibitem[Cortese et al.(2010)]{2010A&A...518L..63C} Cortese, L., Bendo, G.~J., Boselli, A., et al.\ 2010b, \aap, 518, L63 
%\bibitem[Cortese et al.(2010)]{2010MNRAS.403L..26C} Cortese, L., Bendo, G.~J., Isaak, K.~G., Davies, J.~I., \& Kent, B.~R.\ 2010b, \mnras, 403, L26 
\bibitem[Cortese et al.(2012)]{2012A&A...540A..52C} Cortese, L., Ciesla, L., Boselli, A., et al.\ 2012a, \aap, 540, A52 
\bibitem[Cortese et al.(2012)]{2012A&A...544A.101C} Cortese, L., Boissier, S., Boselli, A., et al.\ 2012b, \aap, 544, A101 
%\bibitem[Cortese et al.(2014)]{2014MNRAS.440..942C} Cortese, L., Fritz, J., Bianchi, S., et al.\ 2014, \mnras, 440, 942 
%\bibitem[C{\^o}t{\'e} et al.(2004)]{2004ApJS..153..223C} C{\^o}t{\'e}, P., Blakeslee, J.~P., Ferrarese, L., et al.\ 2004, \apjs, 153, 223 
%\bibitem[C{\^o}t{\'e} et al.(2006)]{2006ApJS..165...57C} C{\^o}t{\'e}, P., Piatek, S., Ferrarese, L., et al.\ 2006, \apjs, 165, 57 
%\bibitem[C{\^o}t{\'e} et al.(2009)]{2009AJ....138.1037C} C{\^o}t{\'e}, S., Draginda, A., Skillman, E.~D., \& Miller, B.~W.\ 2009, \aj, 138, 1037 
%\bibitem[Cowie \& Binney(1977)]{1977ApJ...215..723C} Cowie, L.~L., \& Binney, J.\ 1977, \apj, 215, 723 
\bibitem[Cowie \& Songaila(1977)]{1977Natur.266..501C} Cowie, L.~L., \& Songaila, A.\ 1977, \nat, 266, 501 
%\bibitem[Cowie et al.(1996)]{1996AJ....112..839C} Cowie, L.~L., Songaila, A., Hu, E.~M., \& Cohen, J.~G.\ 1996, \aj, 112, 839 
%\bibitem[Crain et al.(2015)]{2015MNRAS.450.1937C} Crain, R.~A., Schaye, J., Bower, R.~G., et al.\ 2015, \mnras, 450, 1937 
\bibitem[Cramer et al.(2019)]{2019ApJ...870...63C} Cramer, W.~J., Kenney, J.~D.~P., Sun, M., et al.\ 2019, \apj, 870, 63
%\bibitem[Crawford et al.(1999)]{1999MNRAS.306..857C} Crawford, C.~S., Allen, S.~W., Ebeling, H., Edge, A.~C., \& Fabian, A.~C.\ 1999, \mnras, 306, 857 
%\bibitem[Crawford et al.(2005)]{2005MNRAS.363..216C} Crawford, C.~S., Hatch, N.~A., Fabian, A.~C., \& Sanders, J.~S.\ 2005, \mnras, 363, 216 
\bibitem[Crowl \& Kenney(2008)]{2008AJ....136.1623C} Crowl, H.~H., \& Kenney, J.~D.~P.\ 2008, \aj, 136, 1623 
%\bibitem[Crowl et al.(2005)]{2005AJ....130...65C} Crowl, H.~H., Kenney, J.~D.~P., van Gorkom, J.~H., \& Vollmer, B.\ 2005, \aj, 130, 65 
%\bibitem[Cui et al.(2014)]{2014MNRAS.437..816C} Cui, W., Murante, G., Monaco, P., et al.\ 2014, \mnras, 437, 816 
%\bibitem[da Silva et al.(2014)]{2014MNRAS.444.3275D} da Silva, R.~L., Fumagalli, M., \& Krumholz, M.~R.\ 2014, \mnras, 444, 3275 
%\bibitem[David et al.(2014)]{2014ApJ...792...94D} David, L.~P., Lim, J., Forman, W., et al.\ 2014, \apj, 792, 94 
%\bibitem[Davies et al.(2010)]{2010A&A...518L..48D} Davies, J.~I., Baes, M., Bendo, G.~J., et al.\ 2010, \aap, 518, L48 
%\bibitem[Decarli et al.(2007)]{2007MNRAS.381..136D} Decarli, R., Gavazzi, G., Arosio, I., et al.\ 2007, \mnras, 381, 136 
%\bibitem[De Looze et al.(2013)]{2013MNRAS.436.1057D} De Looze, I., Baes, M., Boselli, A., et al.\ 2013, \mnras, 436, 1057 
%\bibitem[De Lucia \& Blaizot(2007)]{2007MNRAS.375....2D} De Lucia, G., \& Blaizot, J.\ 2007, \mnras, 375, 2 
\bibitem[De Lucia et al.(2012)]{2012MNRAS.423.1277D} De Lucia, G., Weinmann, S., Poggianti, B.~M., et al.\ 2012, \mnras, 423, 1277
%\bibitem[DeMaio et al.(2015)]{2015MNRAS.448.1162D} DeMaio, T., Gonzalez, A.~H., Zabludoff, A., Zaritsky, D., \& Brada{\v c}, M.\ 2015, \mnras, 448, 1162 
%\bibitem[Dennison et al.(1998)]{1998PASA...15..147D} Dennison, B., Simonetti, J.~H., \& Topasna, G.~A.\ 1998, \pasa, 15, 147 
%\bibitem[de Vaucouleurs(1961)]{1961ApJS....6..213D} de Vaucouleurs, G.\ 1961, \apjs, 6, 213 
%\bibitem[Dopita \& Sutherland(1995)]{1995ApJ...455..468D} Dopita, M.~A., \& Sutherland, R.~S.\ 1995, \apj, 455, 468 
%\bibitem[Dopita et al.(1997)]{1997ApJ...490..202D} Dopita, M.~A., Koratkar, A.~P., Allen, M.~G., et al.\ 1997, \apj, 490, 202 
\bibitem[Dressler(1980)]{1980ApJ...236..351D} Dressler, A.\ 1980, \apj, 236, 351 
\bibitem[Dressler(2004)]{2004cgpc.symp..206D} Dressler, A.\ 2004, Clusters of Galaxies: Probes of Cosmological Structure and Galaxy Evolution, 206 
\bibitem[Dressler et al.(1997)]{1997ApJ...490..577D} Dressler, A., Oemler, A., Jr., Couch, W.~J., et al.\ 1997, \apj, 490, 577 
%\bibitem[Dressler et al.(1999)]{1999ApJS..122...51D} Dressler, A., Smail, I., Poggianti, B.~M., et al.\ 1999, \apjs, 122, 51 
%\bibitem[Dressler et al.(2013)]{2013ApJ...770...62D} Dressler, A., Oemler, A., Jr., Poggianti, B.~M., et al.\ 2013, \apj, 770, 62 
%\bibitem[Drew et al.(2005)]{2005MNRAS.362..753D} Drew, J.~E., Greimel, R., Irwin, M.~J., et al.\ 2005, \mnras, 362, 753 
%\bibitem[Drew et al.(2014)]{2014MNRAS.440.2036D} Drew, J.~E., Gonzalez-Solares, E., Greimel, R., et al.\ 2014, \mnras, 440, 2036 
%\bibitem[Drissen et al.(2010)]{2010SPIE.7735E..0BD} Drissen, L., Bernier, A.-P., Rousseau-Nepton, L., et al.\ 2010, \procspie, 7735, 77350B 
%\bibitem[Duc et al.(2011)]{2011MNRAS.417..863D} Duc, P.-A., Cuillandre, J.-C., Serra, P., et al.\ 2011, \mnras, 417, 863 
%\bibitem[Duc et al.(2015)]{2015MNRAS.446..120D} Duc, P.-A., Cuillandre, J.-C., Karabal, E., et al.\ 2015, \mnras, 446, 120 
%\bibitem[Duffy et al.(2012)]{2012MNRAS.426.3385D} Duffy, A.~R., Meyer, M.~J., Staveley-Smith, L., et al.\ 2012, \mnras, 426, 3385 
%\bibitem[Durrell et al.(2014)]{2014ApJ...794..103D} Durrell, P.~R., C{\^o}t{\'e}, P., Peng, E.~W., et al.\ 2014, \apj, 794, 103 
%\bibitem[Durret et al.(2011)]{2011A&A...535A..65D} Durret, F., Adami, C., Cappi, A., et al.\ 2011, \aap, 535, A65 
%\bibitem[Dwarakanath et al.(1994)]{1994ApJ...432..469D} Dwarakanath, K.~S., van Gorkom, J.~H., \& Owen, F.~N.\ 1994, \apj, 432, 469 
%\bibitem[Ellison et al.(2008)]{2008AJ....135.1877E} Ellison, S.~L., Patton, D.~R., Simard, L., \& McConnachie, A.~W.\ 2008, \aj, 135, 1877 
%\bibitem[Emsellem et al.(2014)]{2014MNRAS.445L..79E} Emsellem, E., Krajnovi{\'c}, D., \& Sarzi, M.\ 2014, \mnras, 445, L79 
%\bibitem[Evrard et al.(2008)]{2008ApJ...672..122E} Evrard, A.~E., Bialek, J., Busha, M., et al.\ 2008, \apj, 672, 122-137 
%\bibitem[Fabian(1994)]{1994ARA&A..32..277F} Fabian, A.~C.\ 1994, \araa, 32, 277 
%\bibitem[Fabian \& Nulsen(1977)]{1977MNRAS.180..479F} Fabian, A.~C., \& Nulsen, P.~E.~J.\ 1977, \mnras, 180, 479 
%\bibitem[Fanaroff \& Riley(1974)]{1974MNRAS.167P..31F} Fanaroff, B.~L., \& Riley, J.~M.\ 1974, \mnras, 167, 31P 
%\bibitem[Feldmeier et al.(1997)]{1997ApJ...479..231F} Feldmeier, J.~J., Ciardullo, R., \& Jacoby, G.~H.\ 1997, \apj, 479, 231 
%\bibitem[Feldmeier et al.(1998)]{1998ApJ...503..109F} Feldmeier, J.~J., Ciardullo, R., \& Jacoby, G.~H.\ 1998, \apj, 503, 109 
%\bibitem[Feldmeier et al.(2003)]{2003ApJS..145...65F} Feldmeier, J.~J., Ciardullo, R., Jacoby, G.~H., \& Durrell, P.~R.\ 2003, \apjs, 145, 65 
%\bibitem[Ferland et al.(2009)]{2009MNRAS.392.1475F} Ferland, G.~J., Fabian, A.~C., Hatch, N.~A., et al.\ 2009, \mnras, 392, 1475 
%\bibitem[Ferrarese et al.(2006)]{2006ApJS..164..334F} Ferrarese, L., C{\^o}t{\'e}, P., Jord{\'a}n, A., et al.\ 2006, \apjs, 164, 334 
%\bibitem[Ferrarese et al.(2012)]{2012ApJS..200....4F} Ferrarese, L., C{\^o}t{\'e}, P., Cuillandre, J.-C., et al.\ 2012, \apjs, 200, 4 
%\bibitem[Ferrarese et al.(2016)]{2016ApJ...824...10F} Ferrarese, L., C{\^o}t{\'e}, P., S{\'a}nchez-Janssen, R., et al.\ 2016, \apj, 824, 10 
%\bibitem[Fillingham et al.(2015)]{2015MNRAS.454.2039F} Fillingham, S.~P., Cooper, M.~C., Wheeler, C., et al.\ 2015, \mnras, 454, 2039 
%\bibitem[Finkbeiner et al.(2016)]{2016ApJ...822...66F} Finkbeiner, D.~P., Schlafly, E.~F., Schlegel, D.~J., et al.\ 2016, \apj, 822, 66 
%\bibitem[Finoguenov et al.(2008)]{2008ApJ...686..911F} Finoguenov, A., Ruszkowski, M., Jones, C., et al.\ 2008, \apj, 686, 911-917 
%\bibitem[Fitzpatrick(1999)]{1999PASP..111...63F} Fitzpatrick, E.~L.\ 1999, \pasp, 111, 63 
%\bibitem[For et al.(2012)]{2012MNRAS.425.1934F} For, B.-Q., Koribalski, B.~S., \& Jarrett, T.~H.\ 2012, \mnras, 425, 1934 
%\bibitem[Ford \& Butcher(1979)]{1979ApJS...41..147F} Ford, H.~C., \& Butcher, H.\ 1979, \apjs, 41, 147 
%\bibitem[Forman et al.(2007)]{2007ApJ...665.1057F} Forman, W., Jones, C., Churazov, E., et al.\ 2007, \apj, 665, 1057 
%\bibitem[Forman et al.(2017)]{2017ApJ...844..122F} Forman, W., Churazov, E., Jones, C., et al.\ 2017, \apj, 844, 122 
\bibitem[Fossati et al.(2012)]{2012A&A...544A.128F} Fossati, M., Gavazzi, G., Boselli, A., \& Fumagalli, M.\ 2012, \aap, 544, A128 
\bibitem[Fossati et al.(2016)]{2016MNRAS.455.2028F} Fossati, M., Fumagalli, M., Boselli, A., et al.\ 2016, \mnras, 455, 2028 
%\bibitem[Fossati et al.(2017)]{2017ApJ...835..153F} Fossati, M., Wilman, D.~J., Mendel, J.~T., et al.\ 2017a, \apj, 835, 153 
%\bibitem[Fossati et al.(2017)]{} Fossati, M., Mendes, t., Boselli, A., et al., submitted to A\&A (paper III)
\bibitem[Fossati et al.(2018)]{2018A&A...614A..57F} Fossati, M., Mendel, J.~T., Boselli, A., et al.\ 2018, \aap, 614, A57 
%\bibitem[Fouqu{\'e} et al.(2001)]{2001A&A...375..770F} Fouqu{\'e}, P., Solanes, J.~M., Sanchis, T., \& Balkowski, C.\ 2001, \aap, 375, 770 
%\bibitem[Franx \& Illingworth(1990)]{1990ApJ...359L..41F} Franx, M., \& Illingworth, G.\ 1990, \apjl, 359, L41 
%\bibitem[Fryxell et al.(2000)]{2000ApJS..131..273F} Fryxell, B., Olson, K., Ricker, P., et al.\ 2000, \apjs, 131, 273 
%\bibitem[Ftaclas et al.(1984)]{1984ApJ...282...19F} Ftaclas, C., Struble, M.~F., \& Fanelli, M.~N.\ 1984, \apj, 282, 19 
%\bibitem[Fujita(2004)]{2004PASJ...56...29F} Fujita, Y.\ 2004, \pasj, 56, 29 
%\bibitem[Fukazawa et al.(2004)]{2004PASJ...56..965F} Fukazawa, Y., Makishima, K., \& Ohashi, T.\ 2004, \pasj, 56, 965 
%\bibitem[Fukugita et al.(1995)]{1995PASP..107..945F} Fukugita, M., Shimasaku, K., \& Ichikawa, T.\ 1995, \pasp, 107, 945 
\bibitem[Fumagalli et al.(2009)]{2009ApJ...697.1811F} Fumagalli, M., Krumholz, M.~R., Prochaska, J.~X., et al.\ 2009, \apj, 697, 1811
%\bibitem[Fumagalli et al.(2011)]{2011A&A...528A..46F} Fumagalli, M., Gavazzi, G., Scaramella, R., \& Franzetti, P.\ 2011b, \aap, 528, A46 
%\bibitem[Fumagalli et al.(2011)]{2011ApJ...741L..26F} Fumagalli, M., da Silva, R.~L., \& Krumholz, M.~R.\ 2011a, \apjl, 741, L26 
%\bibitem[Fumagalli et al.(2014)]{2014MNRAS.445.4335F} Fumagalli, M., Fossati, M., Hau, G.~K.~T., et al.\ 2014, \mnras, 445, 4335 
%\bibitem[Garel et al.(2015)]{2015MNRAS.450.1279G} Garel, T., Blaizot, J., Guiderdoni, B., et al.\ 2015, \mnras, 450, 1279 
%\bibitem[Garel et al.(2016)]{2016MNRAS.455.3436G} Garel, T., Guiderdoni, B., \& Blaizot, J.\ 2016, \mnras, 455, 3436 
%\bibitem[Gaustad et al.(2001)]{2001PASP..113.1326G} Gaustad, J.~E., McCullough, P.~R., Rosing, W., \& Van Buren, D.\ 2001, \pasp, 113, 1326 
%\bibitem[Gavazzi \& Boselli(1999)]{1999A&A...343...86G} Gavazzi, G., \& Boselli, A.\ 1999, \aap, 343, 86 
%\bibitem[Gavazzi \& Jaffe(1987)]{1987A&A...186L...1G} Gavazzi, G., \& Jaffe, W.\ 1987, \aap, 186, L1 
%\bibitem[Gavazzi et al.(1991)]{1991AJ....101.1207G} Gavazzi, G., Boselli, A., \& Kennicutt, R.\ 1991, \aj, 101, 1207 
%\bibitem[Gavazzi et al.(1996)]{1996A&A...312..397G} Gavazzi, G., Pierini, D., \& Boselli, A.\ 1996, \aap, 312, 397 
%\bibitem[Gavazzi et al.(1998)]{1998AJ....115.1745G} Gavazzi, G., Catinella, B., Carrasco, L., Boselli, A., \& Contursi, A.\ 1998, \aj, 115, 1745 
\bibitem[Gavazzi et al.(1999)]{1999MNRAS.304..595G} Gavazzi, G., Boselli, A., Scodeggio, M., Pierini, D., \& Belsole, E.\ 1999, \mnras, 304, 595 
%\bibitem[Gavazzi et al.(2000)]{2000A&A...361....1G} Gavazzi, G., Boselli, A., V{\'{\i}}lchez, J.~M., Iglesias-Paramo, J., \& Bonfanti, C.\ 2000, \aap, 361, 1 
%\bibitem[Gavazzi et al.(2001)]{2001ApJ...563L..23G} Gavazzi, G., Boselli, A., Mayer, L., et al.\ 2001, \apjl, 563, L23 
%\bibitem[Gavazzi et al.(2002)]{2002A&A...386..114G} Gavazzi, G., Boselli, A., Pedotti, P., Gallazzi, A., \& Carrasco, L.\ 2002, \aap, 386, 114 
%\bibitem[Gavazzi et al.(2002)]{2002A&A...396..449G} Gavazzi, G., Boselli, A., Pedotti, P., Gallazzi, A., \& Carrasco, L.\ 2002a, \aap, 396, 449 
%\bibitem[Gavazzi et al.(2002)]{2002ApJ...576..135G} Gavazzi, G., Bonfanti, C., Sanvito, G., Boselli, A., \& Scodeggio, M.\ 2002b, \apj, 576, 135 
%\bibitem[Gavazzi et al.(2003)]{2003ApJ...597..210G} Gavazzi, G., Cortese, L., Boselli, A., et al.\ 2003a, \apj, 597, 210 
\bibitem[Gavazzi et al.(2003)]{2003A&A...400..451G} Gavazzi, G., Boselli, A., Donati, A., Franzetti, P., \& Scodeggio, M.\ 2003, \aap, 400, 451 
\bibitem[Gavazzi et al.(2004)]{2004A&A...417..499G} Gavazzi, G., Zaccardo, A., Sanvito, G., Boselli, A., \& Bonfanti, C.\ 2004, \aap, 417, 499 
\bibitem[Gavazzi et al.(2005)]{2005A&A...429..439G} Gavazzi, G., Boselli, A., van Driel, W., \& O'Neil, K.\ 2005, \aap, 429, 439 
%\bibitem[Gavazzi et al.(2006)]{2006A&A...449..929G} Gavazzi, G., O'Neil, K., Boselli, A., \& van Driel, W.\ 2006a, \aap, 449, 929 
\bibitem[Gavazzi et al.(2006)]{2006A&A...446..839G} Gavazzi, G., Boselli, A., Cortese, L., et al.\ 2006, \aap, 446, 839 
%\bibitem[Gavazzi et al.(2012)]{2012A&A...545A..16G} Gavazzi, G., Fumagalli, M., Galardo, V., et al.\ 2012, \aap, 545, A16 
\bibitem[Gavazzi et al.(2013)]{2013A&A...553A..89G} Gavazzi, G., Fumagalli, M., Fossati, M., et al.\ 2013, \aap, 553, A89 
\bibitem[Gavazzi et al.(2015)]{2015A&A...580A.116G} Gavazzi, G., Consolandi, G., Dotti, M., et al.\ 2015, \aap, 580, A116 
%\bibitem[Gavazzi et al.(2017)]{2017arXiv170906511G} Gavazzi, G., Consolandi, G., Pedraglio, S., et al.\ 2017, arXiv:1709.06511 
%\bibitem[Gavazzi et al.(2014)]{2014ApJ...785..144G} Gavazzi, R., Marshall, P.~J., Treu, T., \& Sonnenfeld, A.\ 2014, \apj, 785, 144 
%\bibitem[Gavazzi et al.(2018)]{2018A&A...611A..28G} Gavazzi, G., Consolandi, G., Pedraglio, S., et al.\ 2018, \aap, 611, A28 
%\bibitem[Geller \& Huchra(1989)]{1989Sci...246..897G} Geller, M.~J., \& Huchra, J.~P.\ 1989, Science, 246, 897 
%\bibitem[Gendron-Marsolais et al.(2018)]{2018MNRAS.479L..28G} Gendron-Marsolais, M., Hlavacek-Larrondo, J., Martin, T.~B., et al.\ 2018, \mnras, 479, L28 
%\bibitem[Genel et al.(2014)]{2014MNRAS.445..175G} Genel, S., Vogelsberger, M., Springel, V., et al.\ 2014, \mnras, 445, 175 
%\bibitem[Giovanelli et al.(2005)]{2005AJ....130.2598G} Giovanelli, R., Haynes, M.~P., Kent, B.~R., et al.\ 2005, \aj, 130, 2598 
%\bibitem[Girardi et al.(1998)]{1998ApJ...505...74G} Girardi, M., Giuricin, G., Mardirossian, F., Mezzetti, M., \& Boschin, W.\ 1998, \apj, 505, 74 
%\bibitem[Gnedin(2003)]{2003ApJ...589..752G} Gnedin, O.~Y.\ 2003, \apj, 589, 752 
%\bibitem[Gomes et al.(2016)]{2016A&A...588A..68G} Gomes, J.~M., Papaderos, P., Kehrig, C., et al.\ 2016, \aap, 588, A68 
\bibitem[G{\'o}mez et al.(2003)]{2003ApJ...584..210G} G{\'o}mez, P.~L., Nichol, R.~C., Miller, C.~J., et al.\ 2003, \apj, 584, 210 
%\bibitem[Goudfrooij \& de Jong(1995)]{1995A&A...298..784G} Goudfrooij, P., \& de Jong, T.\ 1995, \aap, 298, 784 
%\bibitem[Gu{\'e}rou et al.(2015)]{2015ApJ...804...70G} Gu{\'e}rou, A., Emsellem, E., McDermid, R.~M., et al.\ 2015, \apj, 804, 70 
\bibitem[Gullieuszik et al.(2017)]{2017ApJ...846...27G} Gullieuszik, M., Poggianti, B.~M., Moretti, A., et al.\ 2017, \apj, 846, 27
%\bibitem[Gunawardhana et al.(2013)]{2013MNRAS.433.2764G} Gunawardhana, M.~L.~P., Hopkins, A.~M., Bland-Hawthorn, J., et al.\ 2013, \mnras, 433, 2764 
\bibitem[Gunn \& Gott(1972)]{1972ApJ...176....1G} Gunn, J.~E., \& Gott, J.~R., III 1972, \apj, 176, 1 
%\bibitem[Gwyn(2008)]{2008PASP..120..212G} Gwyn, S.~D.~J.\ 2008, \pasp, 120, 212 
%\bibitem[Haffner et al.(2001)]{2001ApJ...556L..33H} Haffner, L.~M., Reynolds, R.~J., \& Tufte, S.~L.\ 2001, \apjl, 556, L33 
%\bibitem[Hamer et al.(2018)]{2018arXiv180309765H} Hamer, S.~L., Fabian, A.~C., Russell, H.~R., et al.\ 2018, arXiv:1803.09765 
%\bibitem[Hartke et al.(2017)]{2017arXiv170306146H} Hartke, J., Arnaboldi, M., Longobardi, A., et al.\ 2017, arXiv:1703.06146 
\bibitem[Haynes, \& Giovanelli(1984)]{1984AJ.....89..758H} Haynes, M.~P., \& Giovanelli, R.\ 1984, \aj, 89, 758
%\bibitem[Haynes \& Giovanelli(1986)]{1986ApJ...306..466H} Haynes, M.~P., \& Giovanelli, R.\ 1986, \apj, 306, 466 
%\bibitem[Haynes et al.(1984)]{1984ARA&A..22..445H} Haynes, M.~P., Giovanelli, R., \& Chincarini, G.~L.\ 1984, \araa, 22, 445 
%\bibitem[Haynes et al.(2007)]{2007ApJ...665L..19H} Haynes, M.~P., Giovanelli, R., \& Kent, B.~R.\ 2007, \apjl, 665, L19 
%\bibitem[Haynes et al.(2011)]{2011AJ....142..170H} Haynes, M.~P., Giovanelli, R., Martin, A.~M., et al.\ 2011, \aj, 142, 170 
%\bibitem[Heckman et al.(1989)]{1989ApJ...338...48H} Heckman, T.~M., Baum, S.~A., van Breugel, W.~J.~M., \& McCarthy, P.\ 1989, \apj, 338, 48 
%\bibitem[Helfer et al.(2003)]{2003ApJS..145..259H} Helfer, T.~T., Thornley, M.~D., Regan, M.~W., et al.\ 2003, \apjs, 145, 259 
\bibitem[Helmboldt et al.(2005)]{2005ApJ...630..824H} Helmboldt, J.~F., Walterbos, R.~A.~M., Bothun, G.~D., et al.\ 2005, \apj, 630, 824
%\bibitem[Helmboldt et al.(2009)]{2009MNRAS.393..478H} Helmboldt, J.~F., Walterbos, R.~A.~M., Bothun, G.~D., O'Neil, K., \& Oey, M.~S.\ 2009, \mnras, 393, 478 
%\bibitem[Heitsch \& Putman(2009)]{2009ApJ...698.1485H} Heitsch, F., \& Putman, M.~E.\ 2009, \apj, 698, 1485 
%\bibitem[Henriques et al.(2015)]{2015MNRAS.451.2663H} Henriques, B.~M.~B., White, S.~D.~M., Thomas, P.~A., et al.\ 2015, \mnras, 451, 2663 
%\bibitem[Hester et al.(2010)]{2010ApJ...716L..14H} Hester, J.~A., Seibert, M., Neill, J.~D., et al.\ 2010, \apjl, 716, L14 
%\bibitem[Hines et al.(1989)]{1989ApJ...347..713H} Hines, D.~C., Eilek, J.~A., \& Owen, F.~N.\ 1989, \apj, 347, 713 
%\bibitem[Hippelein et al.(2003)]{2003A&A...402...65H} Hippelein, H., Maier, C., Meisenheimer, K., et al.\ 2003, \aap, 402, 65 
%\bibitem[Ho et al.(2014)]{2014MNRAS.444.3894H} Ho, I.-T., Kewley, L.~J., Dopita, M.~A., et al.\ 2014, \mnras, 444, 3894 
%\bibitem[Hodge et al.(1989)]{1989PASP..101...32H} Hodge, P., Lee, M.~G., \& Kennicutt, R.~C., Jr.\ 1989, \pasp, 101, 32 
%\bibitem[Hodge et al.(1999)]{1999PASP..111..685H} Hodge, P.~W., Balsley, J., Wyder, T.~K., \& Skelton, B.~P.\ 1999, \pasp, 111, 685 
%\bibitem[Ilbert et al.(2009)]{2009ApJ...690.1236I} Ilbert, O., Capak, P., Salvato, M., et al.\ 2009, \apj, 690, 1236 
%\bibitem[Ivezi{\'c} et al.(2004)]{2004AN....325..583I} Ivezi{\'c}, {\v Z}., Lupton, R.~H., Schlegel, D., et al.\ 2004, Astronomische Nachrichten, 325, 583 
%\bibitem[J{\'a}chym et al.(2013)]{2013A&A...556A..99J} J{\'a}chym, P., Kenney, J.~D.~P., R{\v z}ui{\v c}ka, A., et al.\ 2013, \aap, 556, A99 
%\bibitem[J{\'a}chym et al.(2014)]{2014ApJ...792...11J} J{\'a}chym, P., Combes, F., Cortese, L., Sun, M., \& Kenney, J.~D.~P.\ 2014, \apj, 792, 11 
%\bibitem[Jacoby et al.(1990)]{1990ApJ...356..332J} Jacoby, G.~H., Ciardullo, R., \& Ford, H.~C.\ 1990, \apj, 356, 332 
%\bibitem[Janowiecki et al.(2015)]{2015ApJ...801...96J} Janowiecki, S., Leisman, L., J{\'o}zsa, G., et al.\ 2015, \apj, 801, 96 
%\bibitem[James et al.(2004)]{2004A&A...414...23J} James, P.~A., Shane, N.~S., Beckman, J.~E., et al.\ 2004, \aap, 414, 23 
%\bibitem[Jones et al.(2002)]{2002ApJ...567L.115J} Jones, C., Forman, W., Vikhlinin, A., et al.\ 2002, \apjl, 567, L115 
%\bibitem[Johnstone et al.(2007)]{2007MNRAS.382.1246J} Johnstone, R.~M., Hatch, N.~A., Ferland, G.~J., et al.\ 2007, \mnras, 382, 1246 
%\bibitem[Karachentsev \& Nasonova(2010)]{2010MNRAS.405.1075K} Karachentsev, I.~D., \& Nasonova, O.~G.\ 2010, \mnras, 405, 1075 
%\bibitem[Karachentsev et al.(2014)]{2014ApJ...782....4K} Karachentsev, I.~D., Tully, R.~B., Wu, P.-F., Shaya, E.~J., \& Dolphin, A.~E.\ 2014, \apj, 782, 4 
%\bibitem[Kauffmann et al.(1993)]{1993MNRAS.264..201K} Kauffmann, G., White, S.~D.~M., \& Guiderdoni, B.\ 1993, \mnras, 264, 201 
%\bibitem[Kauffmann et al.(2003)]{2003MNRAS.346.1055K} Kauffmann, G., Heckman, T.~M., Tremonti, C., et al.\ 2003, \mnras, 346, 1055 
%\bibitem[Kauffmann et al.(2004)]{2004MNRAS.353..713K} Kauffmann, G., White, S.~D.~M., Heckman, T.~M., et al.\ 2004, \mnras, 353, 713 
%\bibitem[Kaufman et al.(1999)]{1999ApJ...527..795K} Kaufman, M.~J., Wolfire, M.~G., Hollenbach, D.~J., \& Luhman, M.~L.\ 1999, \apj, 527, 795 
%\bibitem[Kaviraj et al.(2007)]{2007ApJS..173..619K} Kaviraj, S., Schawinski, K., Devriendt, J.~E.~G., et al.\ 2007, \apjs, 173, 619 
\bibitem[Keel et al.(1985)]{1985AJ.....90..708K} Keel, W.~C., Kennicutt, R.~C., Hummel, E., et al.\ 1985, \aj, 90, 708
%\bibitem[Kenney \& Koopmann(1999)]{1999AJ....117..181K} Kenney, J.~D.~P., \& Koopmann, R.~A.\ 1999, \aj, 117, 181 
%\bibitem[Kenney \& Yale(2002)]{2002ApJ...567..865K} Kenney, J.~D.~P., \& Yale, E.~E.\ 2002, \apj, 567, 865 
%\bibitem[Kenney \& Young(1988)]{1988ApJS...66..261K} Kenney, J.~D., \& Young, J.~S.\ 1988, \apjs, 66, 261 
%\bibitem[Kenney et al.(1995)]{1995ApJ...438..135K} Kenney, J.~D.~P., Rubin, V.~C., Planesas, P., \& Young, J.~S.\ 1995, \apj, 438, 135 
%\bibitem[Kenney et al.(1996)]{1996AJ....111..152K} Kenney, J.~D.~P., Koopmann, R.~A., Rubin, V.~C., \& Young, J.~S.\ 1996, \aj, 111, 152 
%\bibitem[Kenney et al.(2004)]{2004AJ....127.3361K} Kenney, J.~D.~P., van Gorkom, J.~H., \& Vollmer, B.\ 2004, \aj, 127, 3361 
%\bibitem[Kenney et al.(2008)]{2008ApJ...687L..69K} Kenney, J.~D.~P., Tal, T., Crowl, H.~H., Feldmeier, J., \& Jacoby, G.~H.\ 2008, \apjl, 687, L69 
%\bibitem[Kenney et al.(2014)]{2014ApJ...780..119K} Kenney, J.~D.~P., Geha, M., J{\'a}chym, P., et al.\ 2014, \apj, 780, 119 
\bibitem[Kenney et al.(2015)]{2015AJ....150...59K} Kenney, J.~D.~P., Abramson, A., \& Bravo-Alfaro, H.\ 2015, \aj, 150, 59
%\bibitem[Kennicutt(1981)]{1981ApJ...247....9K} Kennicutt, R.~C., Jr.\ 1981, \apj, 247, 9 
%\bibitem[Kennicutt(1983)]{1983AJ.....88..483K} Kennicutt, R.~C., Jr.\ 1983, \aj, 88, 483 
\bibitem[Kennicutt(1998)]{1998ARA&A..36..189K} Kennicutt, R.~C., Jr.\ 1998, \araa, 36, 189 
%\bibitem[Kennicutt \& Hodge(1980)]{1980ApJ...241..573K} Kennicutt, R.~C., \& Hodge, P.~W.\ 1980, \apj, 241, 573 
%\bibitem[Kennicutt \& Kent(1983)]{1983AJ.....88.1094K} Kennicutt, R.~C., Jr., \& Kent, S.~M.\ 1983, \aj, 88, 1094 
%\bibitem[Kennicutt \& Keel(1984)]{1984ApJ...279L...5K} Kennicutt, R.~C., Jr., \& Keel, W.~C.\ 1984, \apjl, 279, L5 
%\bibitem[Kennicutt \& Hodge(1986)]{1986ApJ...306..130K} Kennicutt, R.~C., Jr., \& Hodge, P.~W.\ 1986, \apj, 306, 130 
%\bibitem[Kennicutt \& Evans(2012)]{2012ARA&A..50..531K} Kennicutt, R.~C., \& Evans, N.~J.\ 2012, \araa, 50, 531 
%\bibitem[Kennicutt et al.(1989)]{1989ApJ...337..761K} Kennicutt, R.~C., Jr., Edgar, B.~K., \& Hodge, P.~W.\ 1989, \apj, 337, 761 
\bibitem[Kennicutt et al.(1994)]{1994ApJ...435...22K} Kennicutt, R.~C., Tamblyn, P., \& Congdon, C.~E.\ 1994, \apj, 435, 22
%\bibitem[Kennicutt et al.(2007)]{2007ApJ...671..333K} Kennicutt, R.~C., Jr., Calzetti, D., Walter, F., et al.\ 2007, \apj, 671, 333 
%\bibitem[Kennicutt et al.(2008)]{2008ApJS..178..247K} Kennicutt, R.~C., Jr., Lee, J.~C., Funes, J.~G., et al.\ 2008, \apjs, 178, 247-279 
%\bibitem[Kennicutt et al.(2009)]{2009ApJ...703.1672K} Kennicutt, R.~C., Jr., Hao, C.-N., Calzetti, D., et al.\ 2009, \apj, 703, 1672-1695 
%\bibitem[Kewley et al.(2001)]{2001ApJ...556..121K} Kewley, L.~J., Dopita, M.~A., Sutherland, R.~S., Heisler, C.~A., \& Trevena, J.\ 2001, \apj, 556, 121 
%\bibitem[Kewley et al.(2006)]{2006MNRAS.372..961K} Kewley, L.~J., Groves, B., Kauffmann, G., \& Heckman, T.\ 2006, \mnras, 372, 961 
%\bibitem[Kikuchi et al.(2000)]{2000ApJ...531L..95K} Kikuchi, K., Itoh, C., Kushino, A., et al.\ 2000, \apjl, 531, L95 
\bibitem[Kim et al.(2014)]{2014ApJS..215...22K} Kim, S., Rey, S.-C., Jerjen, H., et al.\ 2014, \apjs, 215, 22 
%\bibitem[Knapen(1998)]{1998MNRAS.297..255K} Knapen, J.~H.\ 1998, \mnras, 297, 255 
\bibitem[Koopmann, \& Kenney(1998)]{1998ApJ...497L..75K} Koopmann, R.~A., \& Kenney, J.~D.~P.\ 1998, \apjl, 497, L75
\bibitem[Koopmann et al.(2001)]{2001ApJS..135..125K} Koopmann, R.~A., Kenney, J.~D.~P., \& Young, J.\ 2001, \apjs, 135, 125 
%\bibitem[Kraft et al.(2011)]{2011ApJ...727...41K} Kraft, R.~P., Forman, W.~R., Jones, C., et al.\ 2011, \apj, 727, 41 
%\bibitem[Kwak \& Shelton(2010)]{2010ApJ...719..523K} Kwak, K., \& Shelton, R.~L.\ 2010, \apj, 719, 523 
%\bibitem[Lagache et al.(2000)]{2000A&A...354..247L} Lagache, G., Haffner, L.~M., Reynolds, R.~J., \& Tufte, S.~L.\ 2000, \aap, 354, 247 
\bibitem[Larson et al.(1980)]{1980ApJ...237..692L} Larson, R.~B., Tinsley, B.~M., \& Caldwell, C.~N.\ 1980, \apj, 237, 692 
\bibitem[Lee et al.(2011)]{2011ApJ...735...75L} Lee, J.~H., Hwang, N., \& Lee, M.~G.\ 2011, \apj, 735, 75 
%\bibitem[Lee-Waddell et al.(2017)]{2017arXiv171009947L} Lee-Waddell, K., Serra, P., Koribalski, B., et al.\ 2017, arXiv:1710.09947 
\bibitem[Lewis et al.(2002)]{2002MNRAS.334..673L} Lewis, I., Balogh, M., De Propris, R., et al.\ 2002, \mnras, 334, 673 
%\bibitem[Licitra et al.(2016)]{2016ApJ...829...44L} Licitra, R., Mei, S., Raichoor, A., et al.\ 2016, \apj, 829, 44 
%\bibitem[Lisenfeld et al.(2016)]{2016A&A...590A..92L} Lisenfeld, U., Braine, J., Duc, P.~A., et al.\ 2016, \aap, 590, A92 
%\bibitem[Lisker et al.(2006)]{2006AJ....132.2432L} Lisker, T., Glatt, K., Westera, P., \& Grebel, E.~K.\ 2006, \aj, 132, 2432 
\bibitem[Liu et al.(2013)]{2013ApJ...772...27L} Liu, G., Calzetti, D., Kennicutt, R.~C., Jr., et al.\ 2013, \apj, 772, 27 
%\bibitem[Longobardi et al.(2013)]{2013A&A...558A..42L} Longobardi, A., Arnaboldi, M., Gerhard, O., et al.\ 2013, \aap, 558, A42
%\bibitem[Longobardi et al.(2015)]{2015A&A...579L...3L} Longobardi, A., Arnaboldi, M., Gerhard, O., \& Mihos, J.~C.\ 2015, \aap, 579, L3 
%\bibitem[Longobardi et al.(2015)]{2015A&A...579A.135L} Longobardi, A., Arnaboldi, M., Gerhard, O., \& Hanuschik, R.\ 2015b, \aap, 579, A135 
%\bibitem[Longobardi et al.(2018)]{} Longobardi, A., Arnaboldi, M., Gerhard, O., Pulsoni C., Soldner-Rembold I., 2018, arXiv:1809.10708v
%\bibitem[Lotz et al.(2004)]{2004AJ....128..163L} Lotz, J.~M., Primack, J., \& Madau, P.\ 2004, \aj, 128, 163 
%\bibitem[Ly et al.(2007)]{2007ApJ...657..738L} Ly, C., Malkan, M.~A., Kashikawa, N., et al.\ 2007, \apj, 657, 738 
%\bibitem[Macchetto et al.(1996)]{1996A&AS..120..463M} Macchetto, F., Pastoriza, M., Caon, N., et al.\ 1996, \aaps, 120, 463 
%\bibitem[Machacek et al.(2004)]{2004ApJ...610..183M} Machacek, M.~E., Jones, C., \& Forman, W.~R.\ 2004, \apj, 610, 183 
%\bibitem[Machacek et al.(2006)]{2006ApJ...648..947M} Machacek, M., Nulsen, P.~E.~J., Jones, C., \& Forman, W.~R.\ 2006a, \apj, 648, 947 
%\bibitem[Machacek et al.(2006)]{2006ApJ...644..155M} Machacek, M., Jones, C., Forman, W.~R., \& Nulsen, P.\ 2006b, \apj, 644, 155 
%\bibitem[Magnier \& Cuillandre(2004)]{2004PASP..116..449M} Magnier, E.~A., \& Cuillandre, J.-C.\ 2004, \pasp, 116, 449 
%\bibitem[Magnier et al.(2013)]{2013ApJS..205...20M} Magnier, E.~A., Schlafly, E., Finkbeiner, D., et al.\ 2013, \apjs, 205, 20 
%\bibitem[Malumuth \& Richstone(1984)]{1984ApJ...276..413M} Malumuth, E.~M., \& Richstone, D.~O.\ 1984, \apj, 276, 413 
%\bibitem[Marcelin et al.(1998)]{1998A&A...338....1M} Marcelin, M., Amram, P., Bartlett, J.~G., Valls-Gabaud, D., \& Blanchard, A.\ 1998, \aap, 338, 1 
%\bibitem[Martin et al.(2005)]{2005ApJ...619L...1M} Martin, D.~C., Fanson, J., Schiminovich, D., et al.\ 2005, \apjl, 619, L1 
%\bibitem[Martin et al.(2018)]{2018MNRAS.473.4130M} Martin, T.~B., Drissen, L., \& Melchior, A.-L.\ 2018, \mnras, 473, 4130 
%\bibitem[Matsumoto et al.(2000)]{2000PASJ...52..153M} Matsumoto, H., Tsuru, T.~G., Fukazawa, Y., Hattori, M., \& Davis, D.~S.\ 2000, \pasj, 52, 153 
%\bibitem[Mattila et al.(2007)]{2007ApJ...654L.131M} Mattila, K., Juvela, M., \& Lehtinen, K.\ 2007, \apjl, 654, L131 
%\bibitem[Mayer et al.(2006)]{2006MNRAS.369.1021M} Mayer, L., Mastropietro, C., Wadsley, J., Stadel, J., \& Moore, B.\ 2006, \mnras, 369, 1021 
%\bibitem[McDonald et al.(2010)]{2010ApJ...721.1262M} McDonald, M., Veilleux, S., Rupke, D.~S.~N., \& Mushotzky, R.\ 2010, \apj, 721, 1262 
%\bibitem[McDonald et al.(2012)]{2012ApJ...746..153M} McDonald, M., Veilleux, S., \& Rupke, D.~S.~N.\ 2012, \apj, 746, 153 
%\bibitem[McGee et al.(2009)]{2009MNRAS.400..937M} McGee, S.~L., Balogh, M.~L., Bower, R.~G., Font, A.~S., \& McCarthy, I.~G.\ 2009, \mnras, 400, 937 
\bibitem[McLaughlin(1999)]{1999ApJ...512L...9M} McLaughlin, D.~E.\ 1999, \apjl, 512, L9 
%\bibitem[M{\'e}ndez et al.(2001)]{2001ApJ...563..135M} M{\'e}ndez, R.~H., Riffeser, A., Kudritzki, R.-P., et al.\ 2001, \apj, 563, 135 
\bibitem[Mei et al.(2007)]{2007ApJ...655..144M} Mei, S., Blakeslee, J.~P., C{\^o}t{\'e}, P., et al.\ 2007, \apj, 655, 144 
\bibitem[Merritt(1983)]{1983ApJ...264...24M} Merritt, D.\ 1983, \apj, 264, 24 
%\bibitem[Merritt(1985)]{1985ApJ...289...18M} Merritt, D.\ 1985, \apj, 289, 18 
%\bibitem[Meyer et al.(2004)]{2004MNRAS.350.1195M} Meyer, M.~J., Zwaan, M.~A., Webster, R.~L., et al.\ 2004, \mnras, 350, 1195 
%\bibitem[Michel-Dansac \& Wozniak(2004)]{2004A&A...421..863M} Michel-Dansac, L., \& Wozniak, H.\ 2004, \aap, 421, 863 
%\bibitem[Michielsen et al.(2004)]{2004MNRAS.353.1293M} Michielsen, D., de Rijcke, S., Zeilinger, W.~W., et al.\ 2004, \mnras, 353, 1293 
%\bibitem[Mihos et al.(2009)]{2009ApJ...698.1879M} Mihos, J.~C., Janowiecki, S., Feldmeier, J.~J., Harding, P., \& Morrison, H.\ 2009, \apj, 698, 1879 
%\bibitem[Mihos et al.(2015)]{2015ApJ...809L..21M} Mihos, J.~C., Durrell, P.~R., Ferrarese, L., et al.\ 2015, \apjl, 809, L21 
%\bibitem[Mihos et al.(2017)]{2017ApJ...834...16M} Mihos, J.~C., Harding, P., Feldmeier, J.~J., et al.\ 2017, \apj, 834, 16 
%\bibitem[Million et al.(2010)]{2010MNRAS.407.2046M} Million, E.~T., Werner, N., Simionescu, A., et al.\ 2010, \mnras, 407, 2046 
%\bibitem[Mitchell et al.(2013)]{2013MNRAS.428.2674M} Mitchell, N.~L., Vorobyov, E.~I., \& Hensler, G.\ 2013, \mnras, 428, 2674 
\bibitem[Moore et al.(1998)]{1998ApJ...495..139M} Moore, B., Lake, G., \& Katz, N.\ 1998, \apj, 495, 139 
%\bibitem[Moss \& Whittle(1993)]{1993ApJ...407L..17M} Moss, C., \& Whittle, M.\ 1993, \apjl, 407, L17 
%\bibitem[Moss \& Whittle(2000)]{2000MNRAS.317..667M} Moss, C., \& Whittle, M.\ 2000, \mnras, 317, 667 
%\bibitem[Mu{\~n}oz et al.(2014)]{2014ApJS..210....4M} Mu{\~n}oz, R.~P., Puzia, T.~H., Lan{\c c}on, A., et al.\ 2014, \apjs, 210, 4 
%\bibitem[Murante et al.(2004)]{2004ApJ...607L..83M} Murante, G., Arnaboldi, M., Gerhard, O., et al.\ 2004, \apjl, 607, L83 
%\bibitem[Murante et al.(2007)]{2007MNRAS.377....2M} Murante, G., Giovalli, M., Gerhard, O., et al.\ 2007, \mnras, 377, 2 
\bibitem[Muzzin et al.(2014)]{2014ApJ...796...65M} Muzzin, A., van der Burg, R.~F.~J., McGee, S.~L., et al.\ 2014, \apj, 796, 65
%\bibitem[Nelson et al.(2015)]{2015A&C....13...12N} Nelson, D., Pillepich, A., Genel, S., et al.\ 2015, Astronomy and Computing, 13, 12 
%\bibitem[Norris et al.(2011)]{2011PASA...28..215N} Norris, R.~P., Hopkins, A.~M., Afonso, J., et al.\ 2011, \pasa, 28, 215 
%\bibitem[Nulsen(1982)]{1982MNRAS.198.1007N} Nulsen, P.~E.~J.\ 1982, \mnras, 198, 1007 
%\bibitem[Nulsen \& Bohringer(1995)]{1995MNRAS.274.1093N} Nulsen, P.~E.~J., \& Bohringer, H.\ 1995, \mnras, 274, 1093 
%\bibitem[O'Connell(1999)]{1999ARA&A..37..603O} O'Connell, R.~W.\ 1999, \araa, 37, 603 
%\bibitem[Olsen et al.(2007)]{2007A&A...461...81O} Olsen, L.~F., Benoist, C., Cappi, A., et al.\ 2007, \aap, 461, 81 
%\bibitem[Osterbrock \& Ferland(2006)]{2006agna.book.....O} Osterbrock, D.~E., \& Ferland, G.~J.\ 2006, Astrophysics of gaseous nebulae and active galactic nuclei, 2nd.~ed.~by D.E.~Osterbrock and G.J.~Ferland.~Sausalito, CA: University Science Books, 2006,  
%\bibitem[Ostriker \& Tremaine(1975)]{1975ApJ...202L.113O} Ostriker, J.~P., \& Tremaine, S.~D.\ 1975, \apjl, 202, L113 
%\bibitem[Ouchi et al.(2008)]{2008ApJS..176..301O} Ouchi, M., Shimasaku, K., Akiyama, M., et al.\ 2008, \apjs, 176, 301-330 
%\bibitem[Owen et al.(2000)]{2000ApJ...543..611O} Owen, F.~N., Eilek, J.~A., \& Kassim, N.~E.\ 2000, \apj, 543, 611 
%\bibitem[Papadopoulou et al.(2016)]{2016MNRAS.460.4513P} Papadopoulou, M., Phillipps, S., \& Young, A.~J.\ 2016, \mnras, 460, 4513 
%\bibitem[Pappalardo et al.(2016)]{2016A&A...589A..11P} Pappalardo, C., Bizzocchi, L., Fritz, J., et al.\ 2016, \aap, 589, A11 
%\bibitem[Parker et al.(2005)]{2005MNRAS.362..689P} Parker, Q.~A., Phillipps, S., Pierce, M.~J., et al.\ 2005, \mnras, 362, 689 
\bibitem[Patton et al.(2011)]{2011MNRAS.412..591P} Patton, D.~R., Ellison, S.~L., Simard, L., McConnachie, A.~W., \& Mendel, J.~T.\ 2011, \mnras, 412, 591 
%\bibitem[Patton et al.(2013)]{2013MNRAS.433L..59P} Patton, D.~R., Torrey, P., Ellison, S.~L., Mendel, J.~T., \& Scudder, J.~M.\ 2013, \mnras, 433, L59 
%\bibitem[Patton et al.(2016)]{2016MNRAS.461.2589P} Patton, D.~R., Qamar, F.~D., Ellison, S.~L., et al.\ 2016, \mnras, 461, 2589 
%\bibitem[Peek et al.(2011)]{2011ApJS..194...20P} Peek, J.~E.~G., Heiles, C., Douglas, K.~A., et al.\ 2011, \apjs, 194, 20 
%\bibitem[Peng et al.(2004)]{2004ApJ...602..685P} Peng, E.~W., Ford, H.~C., \& Freeman, K.~C.\ 2004, \apj, 602, 685 
%\bibitem[Peng et al.(2010)]{2010ApJ...721..193P} Peng, Y.-j., Lilly, S.~J., Kova{\v c}, K., et al.\ 2010, \apj, 721, 193 
%\bibitem[Peng et al.(2012)]{2012ApJ...757....4P} Peng, Y.-j., Lilly, S.~J., Renzini, A., \& Carollo, M.\ 2012, \apj, 757, 4 
%\bibitem[Perlman et al.(2001)]{2001ApJ...561L..51P} Perlman, E.~S., Sparks, W.~B., Radomski, J., et al.\ 2001, \apjl, 561, L51 
%\bibitem[Perlman et al.(2007)]{2007ApJ...663..808P} Perlman, E.~S., Mason, R.~E., Packham, C., et al.\ 2007, \apj, 663, 808 
%\bibitem[Perryman et al.(2001)]{2001A&A...369..339P} Perryman, M.~A.~C., de Boer, K.~S., Gilmore, G., et al.\ 2001, \aap, 369, 339 
%\bibitem[Peterson \& Fabian(2006)]{2006PhR...427....1P} Peterson, J.~R., \& Fabian, A.~C.\ 2006, \physrep, 427, 1 
%\bibitem[Pickles(1998)]{1998PASP..110..863P} Pickles, A.~J.\ 1998, \pasp, 110, 863 
%\bibitem[Planck Collaboration et al.(2014)]{2014A&A...571A...1P} Planck Collaboration, Ade, P.~A.~R., Aghanim, N., et al.\ 2014, \aap, 571, A1 
\bibitem[Pleuss et al.(2000)]{2000A&A...361..913P} Pleuss, P.~O., Heller, C.~H., \& Fricke, K.~J.\ 2000, \aap, 361, 913 
\bibitem[Poggianti et al.(1999)]{1999ApJ...518..576P} Poggianti, B.~M., Smail, I., Dressler, A., et al.\ 1999, \apj, 518, 576
\bibitem[Poggianti et al.(2017)]{2017ApJ...844...48P} Poggianti, B.~M., Moretti, A., Gullieuszik, M., et al.\ 2017, \apj, 844, 48 
\bibitem[Poggianti et al.(2019)]{2019MNRAS.482.4466P} Poggianti, B.~M., Gullieuszik, M., Tonnesen, S., et al.\ 2019, \mnras, 482, 4466
%\bibitem[Powalka et al.(2018)]{2018ApJ...856...84P} Powalka, M., Puzia, T.~H., Lan{\c c}on, A., et al.\ 2018, \apj, 856, 84 
%\bibitem[Privon et al.(2017)]{2017ApJ...846...74P} Privon, G.~C., Stierwalt, S., Patton, D.~R., et al.\ 2017, \apj, 846, 74 
%\bibitem[Proxauf et al.(2014)]{2014A&A...561A..10P} Proxauf, B., {\"O}ttl, S., \& Kimeswenger, S.\ 2014, \aap, 561, A10 
%\bibitem[Puchwein et al.(2010)]{2010MNRAS.406..936P} Puchwein, E., Springel, V., Sijacki, D., \& Dolag, K.\ 2010, \mnras, 406, 936 
%\bibitem[Putman et al.(2003)]{2003ApJ...597..948P} Putman, M.~E., Bland-Hawthorn, J., Veilleux, S., et al.\ 2003, \apj, 597, 948 
%\bibitem[Putman et al.(2012)]{2012ARA&A..50..491P} Putman, M.~E., Peek, J.~E.~G., \& Joung, M.~R.\ 2012, \araa, 50, 491 
%\bibitem[Rafieferantsoa et al.(2015)]{2015MNRAS.453.3980R} Rafieferantsoa, M., Dav{\'e}, R., Angl{\'e}s-Alc{\'a}zar, D., et al.\ 2015, \mnras, 453, 3980 
%\bibitem[Raichoor et al.(2014)]{2014ApJ...797..102R} Raichoor, A., Mei, S., Erben, T., et al.\ 2014, \apj, 797, 102 
%\bibitem[Rand(1992)]{1992AJ....103..815R} Rand, R.~J.\ 1992, \aj, 103, 815 
%\bibitem[Randall et al.(2008)]{2008ApJ...688..208R} Randall, S., Nulsen, P., Forman, W.~R., et al.\ 2008, \apj, 688, 208-223 
%\bibitem[Reynolds(1987)]{1987ApJ...323..553R} Reynolds, R.~J.\ 1987, \apj, 323, 553 
%\bibitem[Reynolds et al.(1998)]{1998PASA...15...14R} Reynolds, R.~J., Tufte, S.~L., Haffner, L.~M., Jaehnig, K., \& Percival, J.~W.\ 1998, \pasa, 15, 14 
%\bibitem[Rich et al.(2011)]{2011ApJ...734...87R} Rich, J.~A., Kewley, L.~J., \& Dopita, M.~A.\ 2011, \apj, 734, 87 
%\bibitem[Rich et al.(2015)]{2015ApJS..221...28R} Rich, J.~A., Kewley, L.~J., \& Dopita, M.~A.\ 2015, \apjs, 221, 28 
%\bibitem[Roediger \& Hensler(2005)]{2005A&A...433..875R} Roediger, E., \& Hensler, G.\ 2005, \aap, 433, 875 
%\bibitem[Roediger \& Br{\"u}ggen(2007)]{2007MNRAS.380.1399R} Roediger, E., \& Br{\"u}ggen, M.\ 2007, \mnras, 380, 1399 
%\bibitem[Roediger \& Br{\"u}ggen(2008)]{2008MNRAS.388L..89R} Roediger, E., \& Br{\"u}ggen, M.\ 2008, \mnras, 388, L89 
%\bibitem[Roediger et al.(2011)]{2011MNRAS.416.1983R} Roediger, J.~C., Courteau, S., McDonald, M., \& MacArthur, L.~A.\ 2011, \mnras, 416, 1983 
%\bibitem[Roediger et al.(2014)]{2014MNRAS.443L.114R} Roediger, E., Br{\"u}ggen, M., Owers, M.~S., Ebeling, H., \& Sun, M.\ 2014, \mnras, 443, L114 
%\bibitem[Roediger et al.(2017)]{2017ApJ...836..120R} Roediger, J.~C., Ferrarese, L., C{\^o}t{\'e}, P., et al.\ 2017, \apj, 836, 120 
%\bibitem[Romanishin(1990)]{1990AJ....100..373R} Romanishin, W.\ 1990, \aj, 100, 373 
%\bibitem[Romanowsky et al.(2012)]{2012ApJ...748...29R} Romanowsky, A.~J., Strader, J., Brodie, J.~P., et al.\ 2012, \apj, 748, 29 
%\bibitem[Rousseau-Nepton et al.(2018)]{2018MNRAS.477.4152R} Rousseau-Nepton, L., Robert, C., Martin, R.~P., Drissen, L., \& Martin, T.\ 2018, \mnras, 477, 4152 
%\bibitem[Rozas et al.(1996)]{1996A&A...307..735R} Rozas, M., Beckman, J.~E., \& Knapen, J.~H.\ 1996, \aap, 307, 735 
%\bibitem[Rudick et al.(2010)]{2010ApJ...720..569R} Rudick, C.~S., Mihos, J.~C., Harding, P., et al.\ 2010, \apj, 720, 569 
%\bibitem[Russeil et al.(2005)]{2005A&A...429..497R} Russeil, D., Adami, C., Amram, P., et al.\ 2005, \aap, 429, 497 
%\bibitem[Russell et al.(2017)]{2017MNRAS.472.4024R} Russell, H.~R., McNamara, B.~R., Fabian, A.~C., et al.\ 2017, \mnras, 472, 4024 
%\bibitem[Sabra et al.(2003)]{2003ApJ...584..164S} Sabra, B.~M., Shields, J.~C., Ho, L.~C., Barth, A.~J., \& Filippenko, A.~V.\ 2003, \apj, 584, 164 
%\bibitem[Sakai et al.(2002)]{2002ApJ...578..842S} Sakai, S., Kennicutt, R.~C., Jr., van der Hulst, J.~M., \& Moss, C.\ 2002, \apj, 578, 842 
%\bibitem[Salom{\'e} \& Combes(2008)]{2008A&A...489..101S} Salom{\'e}, P., \& Combes, F.\ 2008, \aap, 489, 101 
%\bibitem[S{\'a}nchez-Gallego et al.(2012)]{2012MNRAS.422.3208S} S{\'a}nchez-Gallego, J.~R., Knapen, J.~H., Wilson, C.~D., et al.\ 2012, \mnras, 422, 3208 
%\bibitem[S{\'a}nchez-Janssen et al.(2008)]{2008ApJ...679L..77S} S{\'a}nchez-Janssen, R., Aguerri, J.~A.~L., \& Mu{\~n}oz-Tu{\~n}{\'o}n, C.\ 2008, \apjl, 679, L77 
%\bibitem[Sarazin(1986)]{1986RvMP...58....1S} Sarazin, C.~L.\ 1986, Reviews of Modern Physics, 58, 1 
%\bibitem[Sarzi et al.(2006)]{2006MNRAS.366.1151S} Sarzi, M., Falc{\'o}n-Barroso, J., Davies, R.~L., et al.\ 2006, \mnras, 366, 1151 
%\bibitem[Sarzi et al.(2018)]{2018MNRAS.478.4084S} Sarzi, M., Spiniello, C., La Barbera, F., Krajnovi{\'c}, D., \& van den Bosch, R.\ 2018, \mnras, 478, 4084 
%\bibitem[Saul et al.(2012)]{2012ApJ...758...44S} Saul, D.~R., Peek, J.~E.~G., Grcevich, J., et al.\ 2012, \apj, 758, 44 
%\bibitem[Schaye et al.(2015)]{2015MNRAS.446..521S} Schaye, J., Crain, R.~A., Bower, R.~G., et al.\ 2015, \mnras, 446, 521 
%\bibitem[Schellenberger \& Reiprich(2015)]{2015A&A...583L...2S} Schellenberger, G., \& Reiprich, T.~H.\ 2015, \aap, 583, L2 
%\bibitem[Schindler et al.(1999)]{1999A&A...343..420S} Schindler, S., Binggeli, B., \& B{\"o}hringer, H.\ 1999, \aap, 343, 420 
%\bibitem[Schlafly \& Finkbeiner(2011)]{2011ApJ...737..103S} Schlafly, E.~F., \& Finkbeiner, D.~P.\ 2011, \apj, 737, 103 
%\bibitem[Schlegel et al.(1998)]{1998ApJ...500..525S} Schlegel, D.~J., Finkbeiner, D.~P., \& Davis, M.\ 1998, \apj, 500, 525 
%\bibitem[Scott et al.(2012)]{2012MNRAS.419L..19S} Scott, T.~C., Cortese, L., Brinks, E., et al.\ 2012, \mnras, 419, L19 
\bibitem[Scoville et al.(2001)]{2001AJ....122.3017S} Scoville, N.~Z., Polletta, M., Ewald, S., et al.\ 2001, \aj, 122, 3017 
%\bibitem[Scudder et al.(2012)]{2012MNRAS.426..549S} Scudder, J.~M., Ellison, S.~L., Torrey, P., Patton, D.~R., \& Mendel, J.~T.\ 2012, \mnras, 426, 549 
%\bibitem[Seon \& Witt(2012)]{2012ApJ...758..109S} Seon, K.-I., \& Witt, A.~N.\ 2012, \apj, 758, 109 
%\bibitem[Serra et al.(2013)]{2013MNRAS.428..370S} Serra, P., Koribalski, B., Duc, P.-A., et al.\ 2013, \mnras, 428, 370 
%\bibitem[Shelton et al.(2012)]{2012ApJ...751..120S} Shelton, R.~L., Kwak, K., \& Henley, D.~B.\ 2012, \apj, 751, 120 
%\bibitem[Skrutskie et al.(2006)]{2006AJ....131.1163S} Skrutskie, M.~F., Cutri, R.~M., Stiening, R., et al.\ 2006, \aj, 131, 1163 
%\bibitem[Simionescu et al.(2007)]{2007A&A...465..749S} Simionescu, A., B{\"o}hringer, H., Br{\"u}ggen, M., \& Finoguenov, A.\ 2007, \aap, 465, 749 
%\bibitem[Simionescu et al.(2018)]{2018MNRAS.475.3004S} Simionescu, A., Tremblay, G., Werner, N., et al.\ 2018, \mnras, 475, 3004 
%\bibitem[Sivakoff et al.(2007)]{2007ApJ...660.1246S} Sivakoff, G.~R., Jord{\'a}n, A., Sarazin, C.~L., et al.\ 2007, \apj, 660, 1246 
%\bibitem[Sivanandam et al.(2014)]{2014ApJ...796...89S} Sivanandam, S., Rieke, M.~J., \& Rieke, G.~H.\ 2014, \apj, 796, 89 
\bibitem[Solanes et al.(2001)]{2001ApJ...548...97S} Solanes, J.~M., Manrique, A., Garc{\'{\i}}a-G{\'o}mez, C., et al.\ 2001, \apj, 548, 97 
%\bibitem[Solanes et al.(2002)]{2002AJ....124.2440S} Solanes, J.~M., Sanchis, T., Salvador-Sol{\'e}, E., Giovanelli, R., \& Haynes, M.~P.\ 2002, \aj, 124, 2440 
%\bibitem[Solomon et al.(1987)]{1987ApJ...319..730S} Solomon, P.~M., Rivolo, A.~R., Barrett, J., \& Yahil, A.\ 1987, \apj, 319, 730 
%\bibitem[Sommer-Larsen et al.(2005)]{2005MNRAS.357..478S} Sommer-Larsen, J., Romeo, A.~D., \& Portinari, L.\ 2005, \mnras, 357, 478 
%\bibitem[Sorce et al.(2016)]{2016MNRAS.460.2015S} Sorce, J.~G., Gottl{\"o}ber, S., Hoffman, Y., \& Yepes, G.\ 2016, \mnras, 460, 2015 
%\bibitem[Sorgho et al.(2017)]{2017MNRAS.464..530S} Sorgho, A., Hess, K., Carignan, C., \& Oosterloo, T.~A.\ 2017, \mnras, 464, 530 
%\bibitem[Sparks et al.(1993)]{1993ApJ...413..531S} Sparks, W.~B., Ford, H.~C., \& Kinney, A.~L.\ 1993, \apj, 413, 531 
%\bibitem[Sparks et al.(2004)]{2004ApJ...607..294S} Sparks, W.~B., Donahue, M., Jord{\'a}n, A., Ferrarese, L., \& C{\^o}t{\'e}, P.\ 2004, \apj, 607, 294 
%\bibitem[Sparks et al.(2009)]{2009ApJ...704L..20S} Sparks, W.~B., Pringle, J.~E., Donahue, M., et al.\ 2009, \apjl, 704, L20 
%\bibitem[Sparks et al.(2012)]{2012ApJ...750L...5S} Sparks, W.~B., Pringle, J.~E., Carswell, R.~F., et al.\ 2012, \apjl, 750, L5 
\bibitem[Spector et al.(2012)]{2012MNRAS.419.2156S} Spector, O., Finkelman, I., \& Brosch, N.\ 2012, \mnras, 419, 2156 
%\bibitem[Stetson(1987)]{1987PASP...99..191S} Stetson, P.~B.\ 1987, \pasp, 99, 191 
%\bibitem[Stierwalt et al.(2015)]{2015ApJ...805....2S} Stierwalt, S., Besla, G., Patton, D., et al.\ 2015, \apj, 805, 2 
%\bibitem[Strong et al.(1988)]{1988A&A...207....1S} Strong, A.~W., Bloemen, J.~B.~G.~M., Dame, T.~M., et al.\ 1988, \aap, 207, 1 
%\bibitem[Suh et al.(2010)]{2010ApJS..187..374S} Suh, H., Jeong, H., Oh, K., et al.\ 2010, \apjs, 187, 374 
%\bibitem[Sun et al.(2006)]{2006ApJ...637L..81S} Sun, M., Jones, C., Forman, W., et al.\ 2006, \apjl, 637, L81 
\bibitem[Sun et al.(2007)]{2007ApJ...671..190S} Sun, M., Donahue, M., \& Voit, G.~M.\ 2007, \apj, 671, 190 
%\bibitem[Sun et al.(2010)]{2010ApJ...708..946S} Sun, M., Donahue, M., Roediger, E., et al.\ 2010, \apj, 708, 946 
%\bibitem[Takano et al.(1989)]{1989Natur.340..289T} Takano, S., Awaki, H., Koyama, K., Kunieda, H., \& Tawara, Y.\ 1989, \nat, 340, 289 
%\bibitem[Tan et al.(2008)]{2008ApJ...689..775T} Tan, J.~C., Beuther, H., Walter, F., \& Blackman, E.~G.\ 2008, \apj, 689, 775-781 
%\bibitem[Taranu et al.(2014)]{2014MNRAS.440.1934T} Taranu, D.~S., Hudson, M.~J., Balogh, M.~L., et al.\ 2014, \mnras, 440, 1934 
%\bibitem[Taylor \& Webster(2005)]{2005ApJ...634.1067T} Taylor, E.~N., \& Webster, R.~L.\ 2005, \apj, 634, 1067 
%\bibitem[Taylor et al.(2012)]{2012MNRAS.423..787T} Taylor, R., Davies, J.~I., Auld, R., \& Minchin, R.~F.\ 2012, \mnras, 423, 787 
%\bibitem[Temi et al.(2007)]{2007ApJ...660.1215T} Temi, P., Brighenti, F., \& Mathews, W.~G.\ 2007, \apj, 660, 1215 
%\bibitem[Theuns \& Warren(1997)]{1997MNRAS.284L..11T} Theuns, T., \& Warren, S.~J.\ 1997, \mnras, 284, L11 
\bibitem[Thilker et al.(2000)]{2000AJ....120.3070T} Thilker, D.~A., Braun, R., \& Walterbos, R.~A.~M.\ 2000, \aj, 120, 3070
%\bibitem[Thilker et al.(2002)]{2002AJ....124.3118T} Thilker, D.~A., Walterbos, R.~A.~M., Braun, R., \& Hoopes, C.~G.\ 2002, \aj, 124, 3118 
%\bibitem[Toloba et al.(2011)]{2011A&A...526A.114T} Toloba, E., Boselli, A., Cenarro, A.~J., et al.\ 2011, \aap, 526, A114 
%\bibitem[Toloba et al.(2014)]{2014ApJS..215...17T} Toloba, E., Guhathakurta, P., Peletier, R.~F., et al.\ 2014, \apjs, 215, 17 
%\bibitem[Tonnesen \& Bryan(2009)]{2009ApJ...694..789T} Tonnesen, S., \& Bryan, G.~L.\ 2009, \apj, 694, 789 
%\bibitem[Tonnesen \& Bryan(2010)]{2010ApJ...709.1203T} Tonnesen, S., \& Bryan, G.~L.\ 2010, \apj, 709, 1203 
%\bibitem[Tonnesen \& Bryan(2012)]{2012MNRAS.422.1609T} Tonnesen, S., \& Bryan, G.~L.\ 2012, \mnras, 422, 1609 
%\bibitem[Tonnesen \& Stone(2014)]{2014ApJ...795..148T} Tonnesen, S., \& Stone, J.\ 2014, \apj, 795, 148 
%\bibitem[Tonnesen et al.(2011)]{2011ApJ...731...98T} Tonnesen, S., Bryan, G.~L., \& Chen, R.\ 2011, \apj, 731, 98 
%\bibitem[Trinchieri \& di Serego Alighieri(1991)]{1991AJ....101.1647T} Trinchieri, G., \& di Serego Alighieri, S.\ 1991, \aj, 101, 1647 
%\bibitem[Tufte et al.(1998)]{1998ApJ...504..773T} Tufte, S.~L., Reynolds, R.~J., \& Haffner, L.~M.\ 1998, \apj, 504, 773 
%\bibitem[Tufte et al.(2002)]{2002ApJ...572L.153T} Tufte, S.~L., Wilson, J.~D., Madsen, G.~J., Haffner, L.~M., \& Reynolds, R.~J.\ 2002, \apjl, 572, L153 
%\bibitem[Tully \& Shaya(1984)]{1984ApJ...281...31T} Tully, R.~B., \& Shaya, E.~J.\ 1984, \apj, 281, 31 
%\bibitem[Ulrich(1978)]{1978ApJ...221..422U} Ulrich, M.-H.\ 1978, \apj, 221, 422 
%\bibitem[Urban et al.(2011)]{2011MNRAS.414.2101U} Urban, O., Werner, N., Simionescu, A., Allen, S.~W., \& B{\"o}hringer, H.\ 2011, \mnras, 414, 2101 
%\bibitem[Vanden Berk et al.(2001)]{2001AJ....122..549V} Vanden Berk, D.~E., Richards, G.~T., Bauer, A., et al.\ 2001, \aj, 122, 549 
%\bibitem[van Gorkom et al.(1989)]{1989AJ.....97..708V} van Gorkom, J.~H., Knapp, G.~R., Ekers, R.~D., et al.\ 1989, \aj, 97, 708 
%\bibitem[Vazdekis et al.(2010)]{2010MNRAS.404.1639V} Vazdekis, A., S{\'a}nchez-Bl{\'a}zquez, P., Falc{\'o}n-Barroso, J., et al.\ 2010, \mnras, 404, 1639 
%\bibitem[Verdugo et al.(2015)]{2015A&A...582A...6V} Verdugo, C., Combes, F., Dasyra, K., Salom{\'e}, P., \& Braine, J.\ 2015, \aap, 582, A6 
%\bibitem[Vogelsberger et al.(2014)]{2014MNRAS.444.1518V} Vogelsberger, M., Genel, S., Springel, V., et al.\ 2014, \mnras, 444, 1518 
%\bibitem[Vollmer(2003)]{2003A&A...398..525V} Vollmer, B.\ 2003, \aap, 398, 525 
%\bibitem[Vollmer(2009)]{2009A&A...502..427V} Vollmer, B.\ 2009, \aap, 502, 427 
%\bibitem[Vollmer \& Huchtmeier(2007)]{2007A&A...462...93V} Vollmer, B., \& Huchtmeier, W.\ 2007, \aap, 462, 93 
%\bibitem[Vollmer et al.(1999)]{1999A&A...349..411V} Vollmer, B., Cayatte, V., Boselli, A., Balkowski, C., \& Duschl, W.~J.\ 1999, \aap, 349, 411 
%\bibitem[Vollmer et al.(2000)]{2000A&A...364..532V} Vollmer, B., Marcelin, M., Amram, P., et al.\ 2000, \aap, 364, 532 
\bibitem[Vollmer et al.(2001)]{2001ApJ...561..708V} Vollmer, B., Cayatte, V., Balkowski, C., \& Duschl, W.~J.\ 2001, \apj, 561, 708 
%\bibitem[Vollmer(2003)]{2003A&A...398..525V} Vollmer, B.\ 2003, \aap, 398, 525
\bibitem[Vollmer et al.(2004)]{2004A&A...419...35V} Vollmer, B., Balkowski, C., Cayatte, V., van Driel, W., \& Huchtmeier, W.\ 2004, \aap, 419, 35 
%\bibitem[Vollmer et al.(2005)]{2005A&A...439..921V} Vollmer, B., Huchtmeier, W., \& van Driel, W.\ 2005, \aap, 439, 921 
\bibitem[Vollmer et al.(2006)]{2006A&A...453..883V} Vollmer, B., Soida, M., Otmianowska-Mazur, K., et al.\ 2006, \aap, 453, 883 
%\bibitem[Vollmer et al.(2008)]{2008A&A...483...89V} Vollmer, B., Soida, M., Chung, A., et al.\ 2008a, \aap, 483, 89 
%\bibitem[Vollmer et al.(2008)]{2008A&A...491..455V} Vollmer, B., Braine, J., Pappalardo, C., \& Hily-Blant, P.\ 2008b, \aap, 491, 455 
%\bibitem[Vollmer et al.(2009)]{2009A&A...496..669V} Vollmer, B., Soida, M., Chung, A., et al.\ 2009, \aap, 496, 669 
%\bibitem[Vollmer et al.(2012)]{2012A&A...537A.143V} Vollmer, B., Soida, M., Braine, J., et al.\ 2012, \aap, 537, A143 
%\bibitem[Voyer et al.(2014)]{2014A&A...569A.124V} Voyer, E.~N., Boselli, A., Boissier, S., et al.\ 2014, \aap, 569, A124 
%\bibitem[Yagi et al.(2007)]{2007ApJ...660.1209Y} Yagi, M., Komiyama, Y., Yoshida, M., et al.\ 2007, \apj, 660, 1209 
%\bibitem[Yagi et al.(2010)]{2010AJ....140.1814Y} Yagi, M., Yoshida, M., Komiyama, Y., et al.\ 2010, \aj, 140, 1814 
%\bibitem[Yagi et al.(2013)]{2013ApJ...778...91Y} Yagi, M., Gu, L., Fujita, Y., et al.\ 2013, \apj, 778, 91 
%\bibitem[Yagi et al.(2017)]{2017ApJ...839...65Y} Yagi, M., Yoshida, M., Gavazzi, G., et al.\ 2017, \apj, 839, 65 
%\bibitem[Yasuda et al.(1997)]{1997ApJS..108..417Y} Yasuda, N., Fukugita, M., \& Okamura, S.\ 1997, \apjs, 108, 417 
%\bibitem[York et al.(2000)]{2000AJ....120.1579Y} York, D.~G., Adelman, J., Anderson, J.~E., Jr., et al.\ 2000, \aj, 120, 1579 
%\bibitem[Yoshida et al.(2002)]{2002ApJ...567..118Y} Yoshida, M., Yagi, M., Okamura, S., et al.\ 2002, \apj, 567, 118 
%\bibitem[Yoshida et al.(2008)]{2008ApJ...688..918Y} Yoshida, M., Yagi, M., Komiyama, Y., et al.\ 2008, \apj, 688, 918 
%\bibitem[Yoshida et al.(2012)]{2012ApJ...749...43Y} Yoshida, M., Yagi, M., Komiyama, Y., et al.\ 2012, \apj, 749, 43 
%\bibitem[Youngblood \& Hunter(1999)]{1999ApJ...519...55Y} Youngblood, A.~J., \& Hunter, D.~A.\ 1999, \apj, 519, 55 
%\bibitem[Young et al.(1995)]{1995ApJS...98..219Y} Young, J.~S., Xie, S., Tacconi, L., et al.\ 1995, \apjs, 98, 219 
%\bibitem[Young et al.(1996)]{1996AJ....112.1903Y} Young, J.~S., Allen, L., Kenney, J.~D.~P., Lesser, A., \& Rownd, B.\ 1996, \aj, 112, 1903 
%\bibitem[Young et al.(2002)]{2002ApJ...579..560Y} Young, A.~J., Wilson, A.~S., \& Mundell, C.~G.\ 2002, \apj, 579, 560 
%\bibitem[Weil et al.(1997)]{1997ApJ...490..664W} Weil, M.~L., Bland-Hawthorn, J., \& Malin, D.~F.\ 1997, \apj, 490, 664 
%\bibitem[Werner et al.(2010)]{2010MNRAS.407.2063W} Werner, N., Simionescu, A., Million, E.~T., et al.\ 2010, \mnras, 407, 2063 
%\bibitem[Werner et al.(2013)]{2013ApJ...767..153W} Werner, N., Oonk, J.~B.~R., Canning, R.~E.~A., et al.\ 2013, \apj, 767, 153 
%\bibitem[Werner et al.(2014)]{2014MNRAS.439.2291W} Werner, N., Oonk, J.~B.~R., Sun, M., et al.\ 2014, \mnras, 439, 2291 
%\bibitem[Wetzel et al.(2013)]{2013MNRAS.432..336W} Wetzel, A.~R., Tinker, J.~L., Conroy, C., \& van den Bosch, F.~C.\ 2013, \mnras, 432, 336 
%\bibitem[Whitaker et al.(2011)]{2011ApJ...735...86W} Whitaker, K.~E., Labb{\'e}, I., van Dokkum, P.~G., et al.\ 2011, \apj, 735, 86 
%\bibitem[White(1976)]{1976MNRAS.174...19W} White, S.~D.~M.\ 1976, \mnras, 174, 19 
%\bibitem[Wilson 1927]{} Wilson E. B., 1927, J. Am. Stat. Assoc., 22, 209
%\bibitem[Wolf et al.(2003)]{2003A&A...401...73W} Wolf, C., Meisenheimer, K., Rix, H.-W., et al.\ 2003, \aap, 401, 73 
%\bibitem[Wolfire et al.(1995)]{1995ApJ...443..152W} Wolfire, M.~G., Hollenbach, D., McKee, C.~F., Tielens, A.~G.~G.~M., \& Bakes, E.~L.~O.\ 1995, \apj, 443, 152 
%\bibitem[Wood et al.(2010)]{2010ApJ...721.1397W} Wood, K., Hill, A.~S., Joung, M.~R., et al.\ 2010, \apj, 721, 1397 
%\bibitem[Woods et al.(2010)]{2010AJ....139.1857W} Woods, D.~F., Geller, M.~J., Kurtz, M.~J., et al.\ 2010, \aj, 139, 1857 
%\bibitem[Wright et al.(2010)]{2010AJ....140.1868W} Wright, E.~L., Eisenhardt, P.~R.~M., Mainzer, A.~K., et al.\ 2010, \aj, 140, 1868-1881 
%\bibitem[Zhang et al.(2013)]{2013ApJ...777..122Z} Zhang, B., Sun, M., Ji, L., et al.\ 2013, \apj, 777, 122 
%\bibitem[Zhu et al.(2008)]{2008ApJ...686..155Z} Zhu, Y.-N., Wu, H., Cao, C., \& Li, H.-N.\ 2008, \apj, 686, 155-171 
\bibitem[Zibetti et al.(2009)]{2009MNRAS.400.1181Z} Zibetti, S., Charlot, S., \& Rix, H.-W.\ 2009, \mnras, 400, 1181



\end{thebibliography}
\end{document}